\documentclass{optica-article}

\journal{opticajournal} 

\articletype{Research Article}

\usepackage{lineno}
\usepackage{ulem}
\usepackage{xcolor}
\usepackage{natbib}
\setcitestyle{numbers,square}

\begin{document}

\title{Perception of Misalignment States for Sky Survey Telescopes with the Digital Twin and the Deep Neural Networks}

\author{Miao Zhang,\authormark{1} Peng Jia,\authormark{1,3*}  Zhengyang Li\authormark{2**}, Wennan Xiang,\authormark{1}  Jiameng Lv,\authormark{1} Rui Sun,\authormark{1}}

\address{\authormark{1}College of Electronic Information and Optical Engineering, Taiyuan University of Technology, Taiyuan, 030024, China\\
\authormark{2}Nanjing Institute of Astronomical Optics \& Technology, Nanjing, 210094, China\\
\authormark{3}Peng Cheng Lab, Shenzhen, 518066, China}

\email{\authormark{*}robinmartin20@gmail.com\\
\authormark{**}zyli@niaot.ac.cn} 


\begin{abstract*} 
Sky survey telescopes play a critical role in modern astronomy, but misalignment of their optical elements can introduce significant variations in point spread functions, leading to reduced data quality. To address this, we need a method to obtain misalignment states, aiding in the reconstruction of accurate point spread functions for data processing methods or facilitating adjustments of optical components for improved image quality. Since sky survey telescopes consist of many optical elements, they result in a vast array of potential misalignment states, some of which are intricately coupled, posing detection challenges. However, by continuously adjusting the misalignment states of optical elements, we can disentangle coupled states. Based on this principle, we propose a deep neural network to extract misalignment states from continuously varying point spread functions in different field of views. To ensure sufficient and diverse training data, we recommend employing a digital twin to obtain data for neural network training. Additionally, we introduce the state graph to store misalignment data and explore complex relationships between misalignment states and corresponding point spread functions, guiding the generation of training data from experiments. Once trained, the neural network estimates misalignment states from observation data, regardless of the impacts caused by atmospheric turbulence, noise, and limited spatial sampling rates in the detector. The method proposed in this paper could be used to provide prior information for the active optics system and the optical system alignment.
\end{abstract*}

\section{Introduction}
In recent years, there has been a growing trend in the design of optical telescopes, aiming for a wider field of view and a smaller F-number. Although this design improves observation efficiency, it also makes telescopes more susceptible to misalignment of their optical elements. Such misalignment introduces highly variable point spread functions (PSFs), which have a detrimental impact on the detection of faint celestial objects and compromise the accuracy of astrometry and photometry results. To overcome these challenges and ensure the acquisition of high-quality scientific results, it is crucial to develop accurate PSF models. These models are built by utilizing real observation data or telemetry data as a reference to obtain PSFs for different fields of view. The PSF models can then be used for data postprocessing or for actively adjusting the optical elements to enhance the telescopes' observational capabilities.\\

There are two distinct methods for modeling PSFs. Telescopes equipped with adaptive optics systems (AO) are capable of obtaining diffraction-limited images within a small field of view, typically ranging from several arcseconds to several arcminutes. Due to the limited number of stars within this field of view, such PSF modeling algorithms rely on wavefront measurements and deformable mirror commands. The success of PSF modeling algorithms for telescopes with AO systems has been widely acknowledged in various observation tasks \cite{Veran1997,Gendron2006,Jolissaint2012,Martin2016,Wagner2018, Turri2022}. On the other hand, telescopes employed in sky survey projects typically lack AO systems, and PSF variations are primarily caused by atmospheric turbulence, dome seeing and misalignment of optical components. Since these telescopes often conduct long-exposure observations, PSFs resulting from atmospheric turbulence and dome seeing tend to exhibit a symmetry structure and can be adequately described using analytical functions \cite{Moffat1969,Kormendy1973,Stetson1992} or statistical models \cite{Jee2011}. Therefore, the main concern lies in addressing the variations in PSFs arising from the misalignment of optical components.\\

Based on the physical model of the imaging process, misalignment of optical elements results in continuous variations in PSFs across the entire field of view. Several approaches have been proposed to extract misalignment states from patterns of PSFs derived from real observation data. At first, star images, extracted from high-quality images, are utilized to diagnose misalignment and jitter in space telescopes \cite{ma2008diagnosing}. Secondly, employing particular aberration patterns within wide-field telescope images, as discussed in Schechter's work \cite{schechter2011generic}, offers a fast and straightforward approach for estimating misalignment states. Thirdly, the issue of telescope misalignment is addressed by gathering field-related aberrations using double Zernike polynomial decomposition and the least-square fitting method \cite{li2015alignment}. Fourth, particle swarm optimization (PSO) \cite{li2020active} is used with normalized point source sensitivity (PSSn) \cite{an2021alignment} to evaluate real-time active optical system errors. Fifth, an alternative option is to measure the telescope's state using a combination of wavefront sensing systems and an internal metrology system \cite{anqichang_internal_motion_metrology_system}, which could provide effective estimation. Sixth, aligning telescopes by matching the geometric features of star images from different fields of view can be done without the need for wavefront sensors or pointing adjustments \cite{bai2021active}. Lastly, the use of machine learning \cite{wu2022machine} offers a more convenient and high-speed alternative compared to traditional field-dependent aberration methods.\\

By employing these methods, we can subsequently calculate the PSF for different fields of view using physics-based propagation algorithms or deep neural networks \cite{zhang2022method}. In recent years, the deep neural network-based approach seems to perform better compared to traditional methods \cite{jia2020point,jia2021point}. Despite the strong representation ability of deep neural network based algorithms, which enables them to capture the relationship between PSFs and misalignment of optical elements, these algorithms face two major challenges. First, telescopes consist of many optical elements, and misalignment states of these elements introduce a large number of degrees of freedom. This leads to the curse of the dimensionality problem \cite{bellman2015applied}, making it difficult to obtain a sufficient amount of high-quality training data to efficiently construct PSFs. Second, PSFs are typically derived from real observation images, which are subject to electron-photon noise and have low spatial sampling rates, leaving along effects brought by the atmospheric turbulence. Consequently, relying solely on PSFs obtained from a single observation image to estimate misalignment states can result in ambiguities. To address these challenges, this study introduces a novel method for estimating misalignment states, offering two significant advancements.
\begin{itemize}
  \item Instead of trying to establish a direct relationship between PSFs and individual misalignment states, we suggest building functions between them by utilising continuous observation data, which includes several continuous misalignment states. This approach effectively solves the problem of ambiguities.
  \item Through the use of the digital twin model, we sample a range of misalignment states and construct a state graph. This state graph could be used to generate huge amount of data with continuous misalignment states. Additionally, it acts as a valuable tool in exploring the relations between misalignment states and PSFs, which aids in the acquisition of real observation data in practical applications.
\end{itemize}
Building on these two contributions, our approach facilitates the estimation of misalignment states across different optical components within a telescope. The fundamental concept of our method is discussed in Section~\ref{METHODS}. To test its performance, we have devised three scenarios, including a Ritchey-Chrétien telescope with misalignment of the secondary mirror, a Primary Focus Telescope with misalignment of the primary mirror, and a Ritchey-Chrétien telescope with misalignment affecting both the primary and the secondary mirror, as detailed in Section~\ref{result}. These scenarios are subjected to rigorous testing that takes into account atmospheric turbulence, readout noise, and limited spatial sampling rates. The test results demonstrate the practical utility of our approach. Finally, our paper concludes by summarizing key findings and outlining future directions, as presented in Section~\ref{conclusion}.\\

\section{Methods}\label{METHODS}
In recent years, several research groups have proposed machine learning based methods to estimate the misalignment states of optical elements \cite{oteo2013new,jia2020point,liu2020misalignment}. Previous studies have focused on reducing computational complexity by conducting sensitivity analyses and considering only the misalignment states of the most sensitive optical elements. However, innovative designs often incorporate multiple optical components to compensate for aberrations \cite{angel2001lsst,hodapp2004design,ackermann2007large,yuan2012optical}. Consequently, the misalignment states of these optical components become significantly more complex, and considering only the most sensitive elements would not be suitable for practical applications. However, considering misalignment states of all optical elements will bring about a serious problem. To illustrate, let's consider an optical telescope with three different optical elements, each with 5 degrees of freedom. In this case, there would be a total of $5^3=125$ degrees of freedom. If we further divide each degree of freedom into 100 different misalignment states, we would end up with $10^{250}$ unique states. The large number of potential misalignment states makes it challenging to sample in real applications and even more difficult to model the relationships between different misalignment states and their corresponding PSFs. Therefore, we encounter two problems in application of the machine learning based misalignment estimation methods:
\begin{itemize}
\item We need to come up with a feasible method to guide us to sample the misalignment states.
\item We should be able to efficiently separate and detect misalignment states in different optical components.
\end{itemize}
To address the initial challenge, we have leveraged the concept of a digital twin for optical telescopes. This pioneering method involves a highly accurate simulation technique coupled with actual test data, enabling the creation of simulated data that faithfully replicates the optical system's behavior. These meticulously crafted data sets are saved as the state graph and serve as valuable training material for a neural network. As the variations in PSF follow identical principles in both the digital twin and real optical systems, it becomes feasible to employ the neural network, trained with digital twin data, to gain insights into the characteristics of misalignment states in actual telescopes.\\

Given that misalignment states and their corresponding PSFs can be viewed as a sequence with continuous misalignment states, it is advantageous to employ a neural network capable of capturing relationships within a series. Therefore, for the second problem, we utilize a convolutional neural network to extract features of PSFs and a recurrent neural network with long short-term memory (LSTM) to establish connections between continuous misalignment states and their corresponding PSFs. To elaborate, we gather training data using the digital twin and construct the deep neural network to detect misalignment based on PSFs extracted from multiple consecutive observation frames. Further details regarding our methodology will be discussed below.\\

\subsection{The Digital Twin of Optical Telescopes}
\subsubsection{The Concept of the Digital Twin of Telescopes}
The concept of the digital twin has found applications across various domains, including space shuttles and instruments, as evidenced by prior research \cite{grieves2017digital, zhuang2018digital}. In particular, it has been used to replicate the mechanical characteristics of radio telescopes \cite{Li2020fast, bednarz2020digital} and to simulate distortions induced by atmospheric turbulence \cite{jia2022digital}. The development of digital twins of telescopes hinges on three pivotal technologies: high-fidelity simulation methods, machine learning-driven state transformation algorithms, and automated measure and adjust instruments.\\

Regarding the simulation method, many different approaches can be employed to generate highly accurate simulated images based on the telescope states at reasonable time and resource costs \cite{wang2012computer, rigaut2013simulating, conan2014object, peterson2015simulation, perrin2016poppy, reeves2016soapy, por2018high, basden2018durham, ferreira2018compass}. The state transformation algorithm establishes relationships between real observation data and simulated data, allowing the introduction of certain hard-to-model effects (e.g., atmospheric turbulence or thermal deformation) and producing high-fidelity simulation data. The advancement of deep neural networks enables the creation of complex functions that generate high-fidelity simulated images based on real observation data \cite{ren2020thermal,han2021dnn,jia2022digital}. Finally, measurement and adjustment instruments, such as automatic hexapods or 4D interferometers, offer improved performance and affordability, facilitating the collection of a large amount of data automatically for building the digital twin.\\

In this study, we employ ZEMAX OpticStudio 22.2\footnote{Supported by the ZEMAX Global Academic Programme.} as our simulator for generating PSFs across different field of views. To streamline our data collection process, we employ a Python script based on pyzdde to automate ZEMAX runs, producing various combinations of optical element misalignment states and their corresponding PSFs. To simulate real-world observational conditions, we introduce atmospheric turbulence using the Moffat model \cite{trujillo2001effects} and the dome seeing with a thin layer of atmospheric turbulence \citep{jia2015simulation}. Besides, we also simulate the atmosphere dispersion. However, we assume the atmosphere dispersion could be effectively reduced by the atmosphere dispersion corrector and we assume there are around 0.4 arcsec residual dispersion \citep{su2012atmospheric}. Lastly, we consider the spatial sampling rate of the detector, ensuring that the generated PSFs matched the appropriate spatial sampling rate. These PSFs, coupled with their corresponding misalignment states, form the data set used to train and test our neural network. It's important to emphasize that this paper centers on algorithm design. Given that gathering a substantial volume of real observation data demands significant human intervention, there's a crucial need to develop an efficient sampling strategy and utilize automated devices for data collection. These designs and findings from laboratory tests will be addressed in our upcoming paper. \\

\subsubsection{The State Graph of Telescopes With Different Misalignment States}
\begin{figure}[htbp]
\centering\includegraphics[width=7cm]{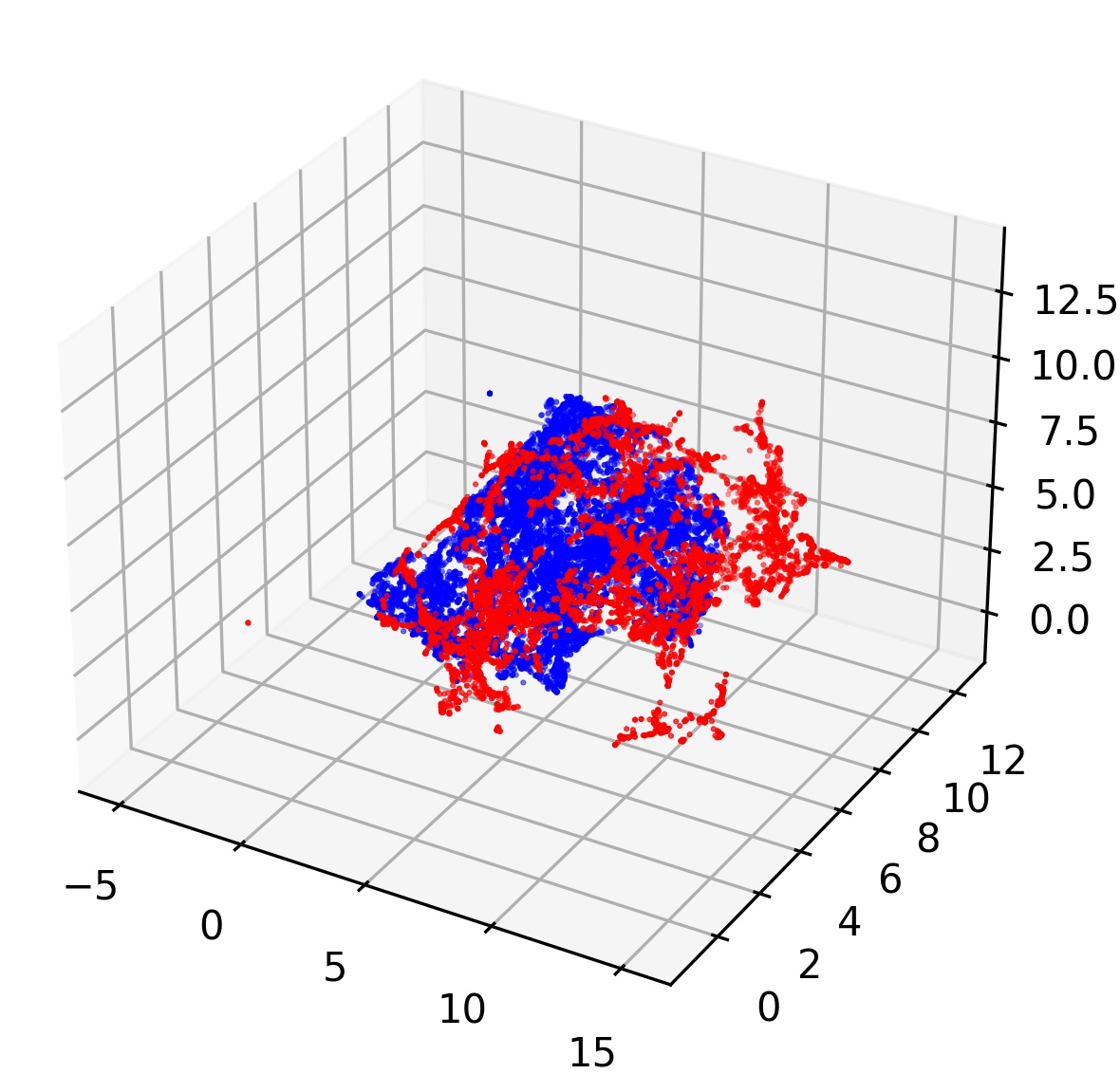}
\caption{The image illustrates a state diagram representing four types of misalignment (x-decenter, y-decenter, x-tilt, y-tilt) in an RC telescope, following dimensionality reduction with UMAP \cite{mcinnes2018umap}. The blue dots represent data points used during the neural network training phase, while the red dots correspond to data points used for testing the model generated during the initial training. By employing an active learning strategy, our model transitions into the space of the red dots, ultimately yielding our final model.}
\label{fig:NewStategraph}
\end{figure}

Since multiple optical elements within a telescope leads to a substantial number of potential misalignment states, it will introduce significant complexity when considering sequences of these states. To tackle this challenge, we propose the utilization of a state graph for the comprehensive modeling and storage of diverse misalignment state data obtained by the digital twin. The schematic representation of the state graph is depicted in Figure \ref{fig:NewStategraph}.\\

\begin{figure}[htbp]
\centering\includegraphics[width=7cm]{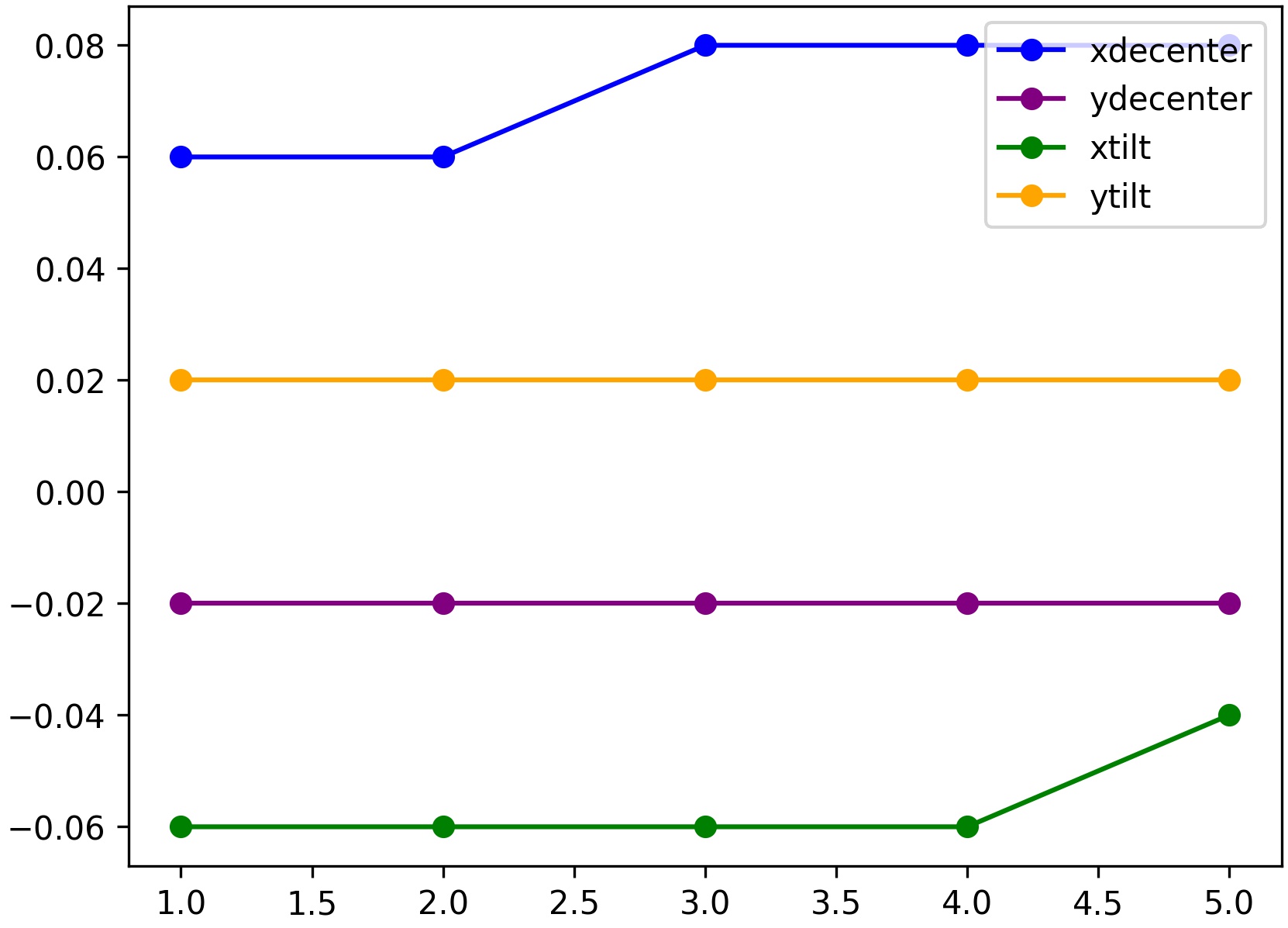}
\caption{We have selected a sequence comprising 5 misalignment states. Along the x-axis, we depict various misalignment states, while the y-axis represents the misalignment values. Each group of points sharing the same x-coordinate corresponds to the misalignment value of a particular state. As illustrated, these misalignment states have resulted in adjustments to x-decenter (state 3) and x-tilt (state 4), while y-decenter and y-tilt remain consistently unchanged throughout the sequence.}
\label{fig:one sequence display}
\end{figure}

The state graph is a type of graph which contains point in N dimensions to represent the state of the telescope. As shown in Figure~\ref{fig:NewStategraph}, the state graph exhibits the following characteristics:
\begin{itemize}
    \item Each point in the state graph stands for a particular misalignment state of the telescope and is associated with PSF maps of the full field of view.
    \item Misalignment states change continuously between adjacent points in the state graph.
    \item A sequence of misalignment states consists of several adjacent points in the state graph.
\end{itemize}
Utilizing the state graph, we can efficiently store and retrieve sequences of misalignment states for the neural network. An illustrative example of such a misalignment sequence is displayed in Figure~\ref{fig:one sequence display}. In the case of a lens with four misalignment dimensions, we have shown a sequence with the length of 5. Within each state, the decision to adjust one misalignment or retain the current configuration is determined randomly. It's worth noting that in this particular sequence, the x-decenter (decenter along the x-direction) shifts in misalignment state 3, while the x-tilt (tilt along the x-direction) varies in misalignment state 5, whereas other misalignment directions remain unchanged.\\

Furthermore, when utilizing the neural network to process data collected by ground-based telescopes, it becomes essential to consider the impact of atmospheric turbulence and the spatial sampling rate of the detector. Consequently, the PSFs obtained through the digital twin require additional processing. To meet this requirement, we store spatial sampling rates and PSFs in hdf5 files \cite{ambatipudi2023comparison}. The hdf5 file keys correspond to specific misalignment states, while the data consists of two-dimensional matrix that represent the full-field PSF and float numbers that represents the spatial sampling rate.\\

\subsubsection{The Data Acquisition Method}
The data generated by the digital twin is utilised to sample the space formed by misalignment states and their corresponding PSFs. We define the range of misalignment states according to manufacturing and assembly tolerances and adjust the minimum sampling step by the optical element's moving part abilities (such as the minimal moving distance of the hexapods). As we gather PSFs from different field of views in each state, it's crucial to adopt an appropriate method to select PSFs from different field of views. Given that both telescopes discussed in this paper are axial symmetry, the distribution of PSFs with varying shapes is theoretically also axially symmetric, provided that misalignment remains within reasonable limits. Furthermore, the PSF experiences gradual alterations near the center of the field of view but displays rapid changes towards the corners of the field of view. After careful consideration, we have opted to employ the ring sampling method, utilizing six rings and eight arms. This choice ensures that each sample point represents an equal area, similar to the method we employed in a previous study \cite{jia2018ground}.\\ 

We use the following steps to obtain PSFs. First, we obtain the optical design of a specific telescope and add coordinate break surfaces in front of and behind the optical elements to introduce misalignment states to the optical system. Next, we create a sampling matrix based on the minimal sampling steps, the range of misalignment, and the number of elements with misalignment. After that, we implement the misalignment states and obtain PSFs for different fields of view corresponding to different misalignment states. In this study, we have chosen the Huygens method to calculate PSFs. We have set the spatial sampling rate to $32\times 32$ pixels for each PSF. All PSFs are stored in the PSF map as a matrix that arranges each PSF row by row and column by column \cite{jia2021point,hughes2023coma}. However, since we obtain PSFs of different fields of view through rings and arms, we use following these steps to organise these PSFs into the PSF map:
\begin{itemize}
\item Resize all PSFs to the same pixel scale and cut them to the same shape.
\item Concatenate PSFs of the same ring in a row and PSFs of the same arm in the same column.
\end{itemize}
By implementing the above steps, we can effectively obtain misalignment states and their corresponding PSF maps that can be further used to train the neural network, as shown in the left bottom of Figure~\ref{fig:seq2seq}.\\

\subsection{Perception of Telescope Misalignment States With the Deep Neural Network}
\subsubsection{The Structure of the Deep Neural Network}
In this study, we utilize sequences of telescope misalignment states along with their associated PSFs to estimate the misalignment states. Therefore, it's advantageous to employ neural networks, which are capable to model sequences. One widely adopted choice for this task is the sequence-to-sequence model, commonly applied in various applications such as machine translation, image captioning, and speech recognition. We feed a sequence of PSF maps as input and obtain an output sequence of corresponding misalignment states. During the training process, the input shape is sequence number $\times$ width of the PSF map $\times$ length of the PSF map, while the output shape is sequence number $\times$ misalignment states. Misalignment states are determined according to our specific requirements and can be calculated by  number of optical components $\times$ number of degrees of freedom for each optical element.\\

\begin{figure}[htbp]
\centering\includegraphics[width=7cm]{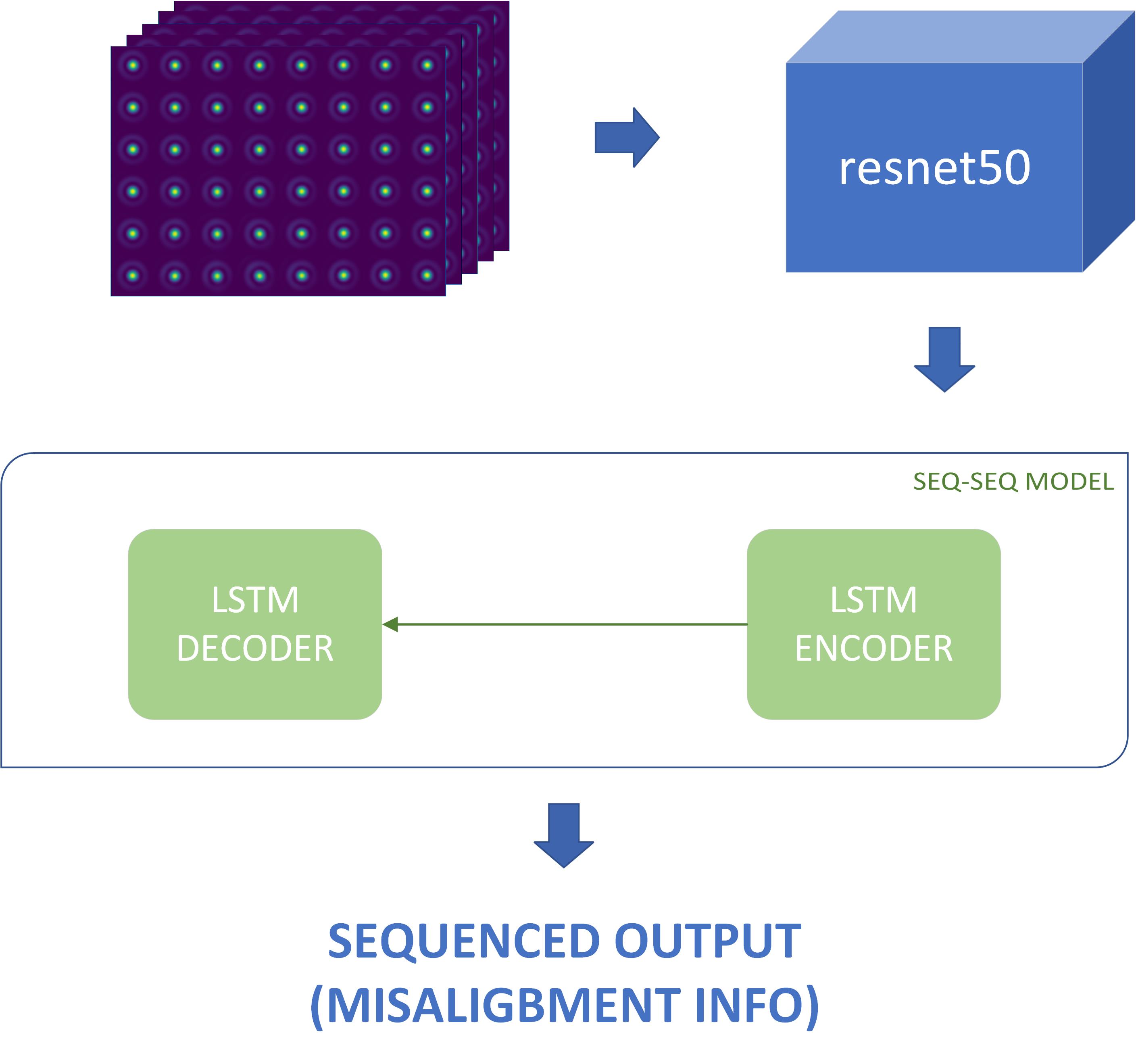}
\caption{The flowchart of the neural network employed in this paper consists of the following components. The input to the neural network is a PSF map, which contains PSFs from different field of views. This PSF map is fed into the ResNet50 to extract relevant features. Subsequently, these features are passed through an encoder and then to a decoder to obtain the corresponding misalignment states for each PSF map.}
\label{fig:seq2seq}
\end{figure}

The model used in this paper, MPNN-Misalignment Perception Neural Network, is basically a sequence-to-sequence model which consists of an encoder and a decoder, as depicted in Figure~\ref{fig:seq2seq}. Initially, ResNet50, a 50-layer Residual Network, is utilized in the encoder to extract PSF features \cite{he2016deep}. To account for the spatial variations of PSFs, we adapt the PSF map to represent PSFs with different field of views, as shown in the upper-left figure of Figure~\ref{fig:seq2seq}. ResNet50 incorporates a distinctive convolutional block structure called the bottleneck, enhancing computational efficiency and reducing parameters. Additionally, ResNet50 use shortcut connections to bypass specific layers and establish direct information flow, thereby improving training efficiency and stability. The extracted features are then passed to subsequent layers, which model relations between sequences.\\

There are several different methods for modeling relationships between sequences, which include Recurrent Neural Networks (RNNs), Long Short-Term Memory networks (LSTMs), and Transformers. While RNNs are capable of handling sequence data, they have limitations in efficiently processing long-range dependencies and can be susceptible to issues like vanishing or exploding gradients \cite{ELMAN1990179}. LSTMs address these limitations by incorporating forget gates and memory gates \cite{10.1162/neco.1997.9.8.1735}. Transformers, originally developed for natural language processing, leverage attention mechanisms to model relationships between components, regardless of their distance from each other \cite{vaswani2017attention}. For our current task, where the sequence is not overly long, LSTM meets our requirements effectively. However, when dealing with more extensive sequences, the Transformer architecture becomes necessary for accurate state recognition\\ 

\begin{figure}[htbp]
\centering\includegraphics[width=9cm]{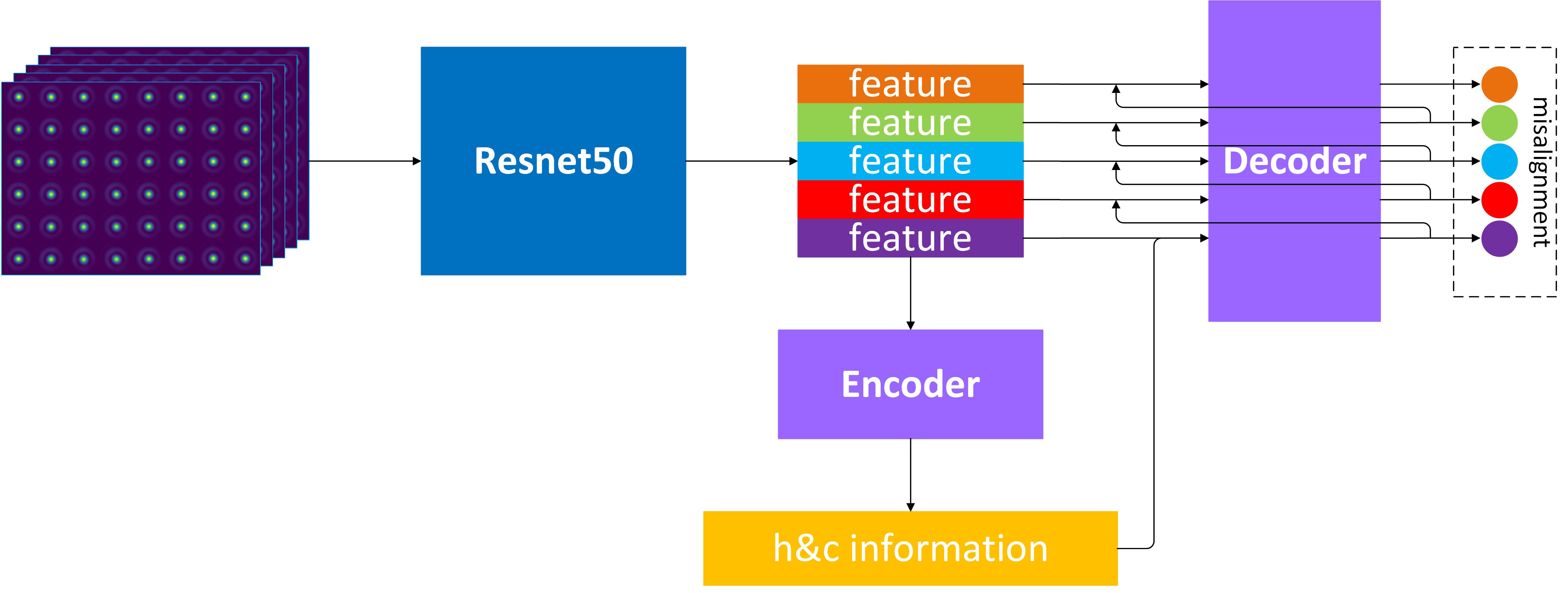}
\caption{The figure shows both the encoder and the decoder, both of which have an identical structure. The encoder receives input features from ResNet50 and chooses the essential features required for estimating misalignment states. Conversely, the decoder deduces misalignment states based on these selected features.}
\label{fig:detailedmodelstruture}
\end{figure}

The encoder and decoder share the same core structure in the LSTM, as shown in Figure~\ref{fig:detailedmodelstruture}. To derive the misalignment states of optical elements as the output of the neural network, we introduce a fully connected layer with a size corresponding to the number of misalignment states. These misalignment states include decenter and tilt along various directions for different optical elements. During the training phase, we calculate the Mean Square Error (MSE) between the output and the target values using Equation~\ref{eq1} as the loss function. The MSE is then used to update the neural network's weights through backpropagation, enabling it to make precise predictions for a variety of input PSF maps.
\begin{equation}
\label{eq1}
    MSE = \frac{1}{N}  \sum_{i=1}^n(prediction_i-target_i)^2.
\end{equation}
In summary, we have adopted the sequence-to-sequence model with the LSTM network to capture the relationship between PSFs and their corresponding misalignment states. For optical systems with varying numbers of optical elements, we can easily adjust the input and output sizes of this model to accommodate specific misalignment scenarios.\\

\subsubsection{The Training Strategy of the Neural Network}
\label{Training}
\begin{figure}[htbp]
\centering\includegraphics[width=9cm]{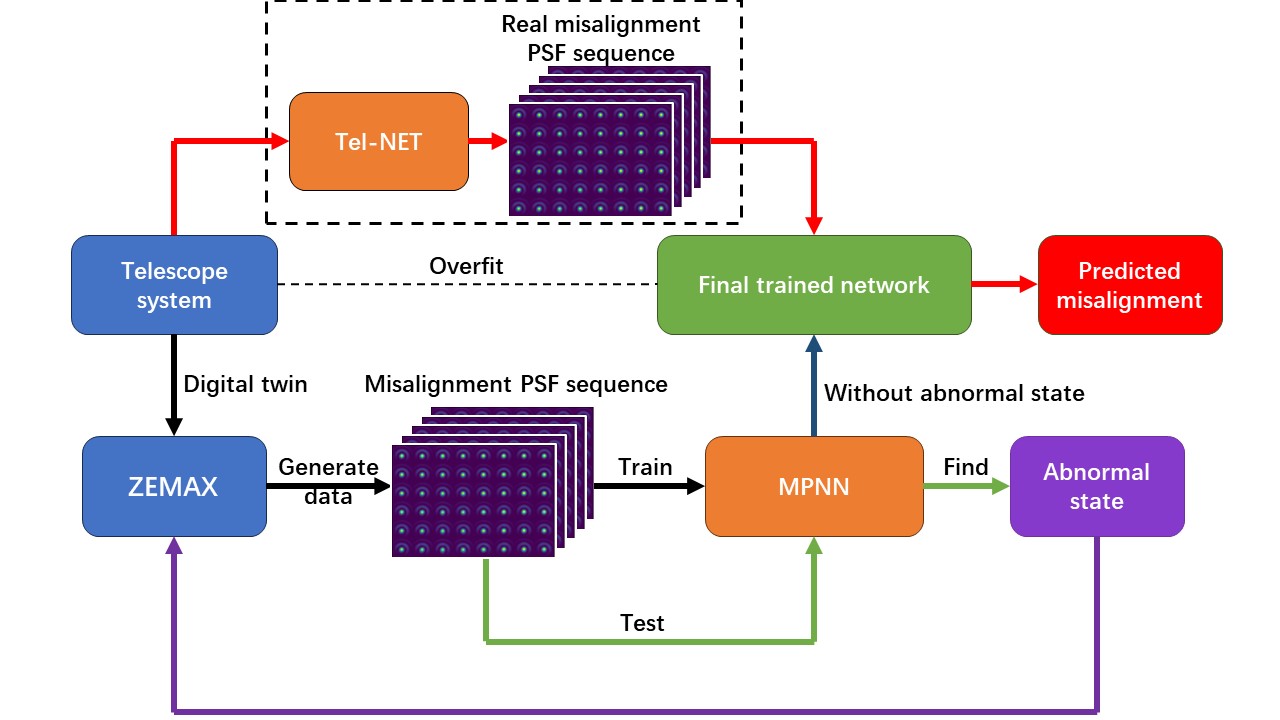}
\caption{Overall training strategy of our network MPNN.}
\label{fig:training strategy}
\end{figure}

The approach presented in this paper diverges from the typical philosophy seen in other machine learning based methods. Leveraging the digital twin, we need to systematically explore all conceivable misalignment states for a telescope. This comprehensive dataset enables the neural network to learn the intricate relationship between misalignment states and their corresponding PSF maps. Our objective is to deliberately over-fit the neural network with misalignment data tailored to a specific telescope. To attain this objective, we utilize the state graph as a guide for generating training data and implement an active learning strategy to enhance the training process of the neural network. We have conducted our experiments using a computer equipped with 8 Nvidia V100 GPU cards, each having 80 GB of memory. There are two reasons behind our choice of this hardware. First and foremost, our approach involves feeding sequences of images as input to the neural network, which requires a larger memory capacity compared to conventional neural networks. Additionally, our active learning strategy involves training with both challenging-to-fit misalignment states and randomly generated misalignment states, necessitating a substantial amount of data to be loaded into the GPU memory. Nevertheless, it's worth noting that such high-end hardware is not always mandatory. The neural networks used in our experiments for RC and PFC telescopes have sizes of 190 MB and 490 MB, respectively, when saved as weight and structure files. Commercial GPUs, such as RTX 3090 or RTX 4090, are also capable of handling this task efficiently. The training procedure is structured as shown in Figure~\ref{fig:training strategy}. Details are shown below below:
\begin{itemize}
\item We initiate the process by randomly selecting a point within the state graph.
\item Subsequently, this point may transition to neighboring points within the state graph or remain stationary.
\item To construct a sequence, we employ the above steps to sample several adjacent points based on the desired length of the state sequence.
\item After generating one sequence, we repeat the aforementioned steps to create multiple sequences.
\item The PSF maps contained within these sequences undergo convolution with PSFs generated by varying atmospheric turbulence coherent lengths (if required) and are down-sampled to align with the spatial sampling rate of the detector.
\item Both the PSF maps and their corresponding misalignment states from these sequences serve as training data for the neural network.
\item We utilize the Adam optimizer with a learning rate of 2e-5 during training.
\item The trained neural network is used to predict misalignment states from PSF maps from the test data set, and we can find that some misalignment states that are hard to predict.
\item In response, we actively generate additional sequences, encompassing these difficult-to-predict states, along with some randomly generated sequences.
\item We iterate through these steps multiple times to further train the neural network.
\end{itemize}

As mentioned earlier, certain misalignment states exhibit a degree of coupling, resulting in similar PSF maps that pose challenges for the neural network to effectively distinguish. Consequently, even after completion of training the neural network with the aforementioned steps, accurately modeling some misalignment states remains a formidable task. With above steps, we could highlight the neural network's limitations. In real applications, two distinct scenarios may unfold. In the first scenario, where sufficient time and appropriate equipment are available during the commissioning phase, we have the opportunity to gather data that include PSF maps along with their corresponding misalignment states. These data can subsequently be utilised to fine-tune the neural network for practical applications. The second scenario occurs when we cannot access the data with absolute calibration. Instead, we can only obtain limited PSF maps and misalignment states. In such a situation, we proceed to train the neural network using all available data while disregarding certain states that may pose challenges for the neural network to fit accurately. Additionally, given the substantial risk associated with the inability to locate stars in predefined positions due to chance, we recommend employing the Tel-Net method outlined in our prior paper \cite{jia2021point} to initially acquire PSFs in these predefined positions. Subsequently, these PSFs can be used for the purpose of estimating misalignment. The complete schematic is depicted in Figure~\ref{fig:training strategy}.\\

\section{Performance Evaluation of the Method} \label{result}
In this section, we will test the performance of our method using two telescopes. The first is a Ritchey-Chrétien telescope (RC), while the second is a primary focus telescope (PFC). To account for real-world application demands, we introduce the influence of atmospheric turbulence, the dome seeing, the residual atmosphere dispersion and spatial sampling effects to the PSFs obtained by these telescopes. This section will delve into the examination of these three scenarios. In the first and second scenarios, we will investigate the impact of secondary mirror misalignment on the RC telescope and primary mirror misalignment on the PFC telescope, respectively. In the third scenario, we will explore the scenario in which both the primary and secondary mirrors of the RC telescope have misalignment states. Detailed discussions of these three scenarios will be presented in the following subsections.\\

\subsection{Scenario I: Misalignment of the Secondary Mirror in a RC Telescope}
\label{Per1}
In this scenario, we will conduct an assessment of our algorithm's performance in estimating the misalignment states of the secondary mirror within a RC telescope. The RC telescope includes two mirrors and a lens corrector, which facilitate moderate image qualities within a medium-sized field of view. The detailed parameters of the RC telescope are described in Table~\ref{tab:RCparameters}. In practical applications, scientists adjust the secondary mirror of the RC telescope to achieve high-quality images. In this context, we will evaluate the capabilities and constraints of our method by performing similar procedures, both with and without the influence of atmospheric turbulence the dome seeing, the residual atmosphere dispersion and the sub-pixel shift.\\

\subsubsection{Misalignment of the Secondary Mirror in a RC Telescope with four misalignment dimensions}
We first will dive into four distinct dimensions of misalignment: tilt and decenter along the x and y directions. For the decenter, we set the minimal step size to $0.01 mm$, with a misalignment range of $\pm 0.1 mm$. For the tilt, the minimal step size is defined as $0.02$ degrees, with a misalignment range of $\pm 0.1$ degrees.\\ 

\begin{table}[htbp]
\centering
\caption{\bf \centering Parameters of the RC telescope in mm}
\begin{tabular}{ccccc}
\hline
Surface&Radius&Thickness&Semi-Diameter&Glass \\
\hline
1&-742.857&-260&78&$Mirror$\\
2&-290.233&471.717&22.566&$Mirror$ \\
3&-55.230&7.5&18&$SF11$\\
4&-118.498&5&18\\
$5(Image)$&$Infinity$&-&0 \\
\hline
\end{tabular}
  \label{tab:RCparameters}
\end{table}

\begin{figure}[htbp]
\centering\includegraphics[width=9cm]{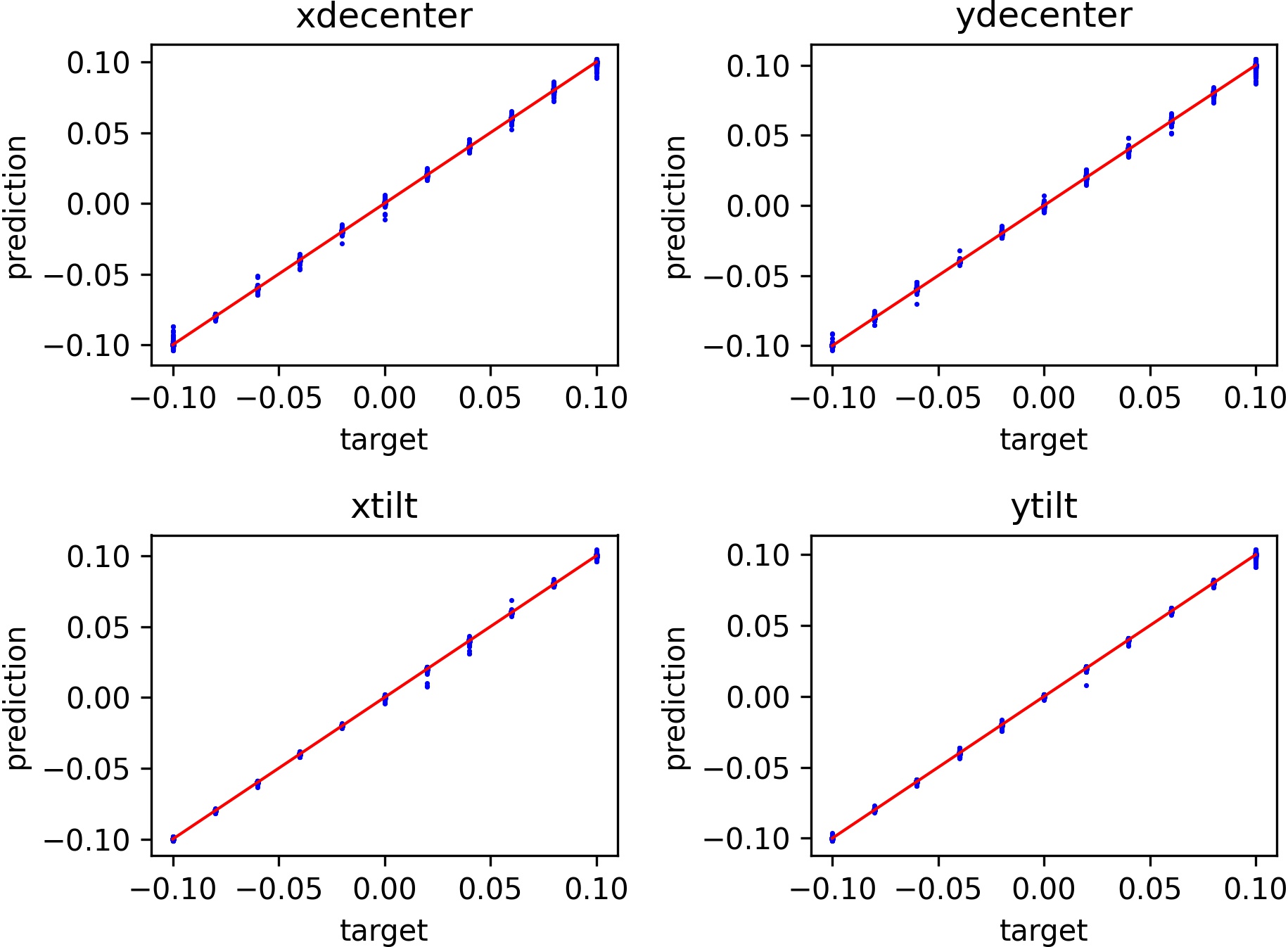}
\caption{Misalignment estimation results for the R-C telescope with 4 misalignment(x-decenter, y-decenter, x-tilt, y-tilt).}
\label{fig:LPRC_final0.02movesresult}
\end{figure}

Using the digital twin, we have obtained a dataset of 5000 suquences which involve around 8000 misalignment states along with their corresponding PSF maps. In our training process, we employed a learning rate of 2e-5 and utilized the Adam optimizer. It's worth noting that within just 100 epochs, the network typically converges. However, owing to the requirement for exceptionally high accuracy, the neural network has been trained by nearly 2000 epochs originally using a random batch of data, then 1000 epochs with the active learning strategy, as elucidated in Section~\ref{Training}. Overall, it takes around 150 hours to train the neural network. Through the initial training of the neural network with a batch of sequences with length of 5 (5 continuous misalignment states), we select another batch randomly to evaluate the model. We conduct this evaluation to explore the neural network's limitations and store sequences with MSE exceeding 1e-6 as outliers. The test results are presented in Figure~\ref{fig:LPRC_final0.02movesresult}.\\

\begin{figure}[htbp]
\centering\includegraphics[width=9cm]{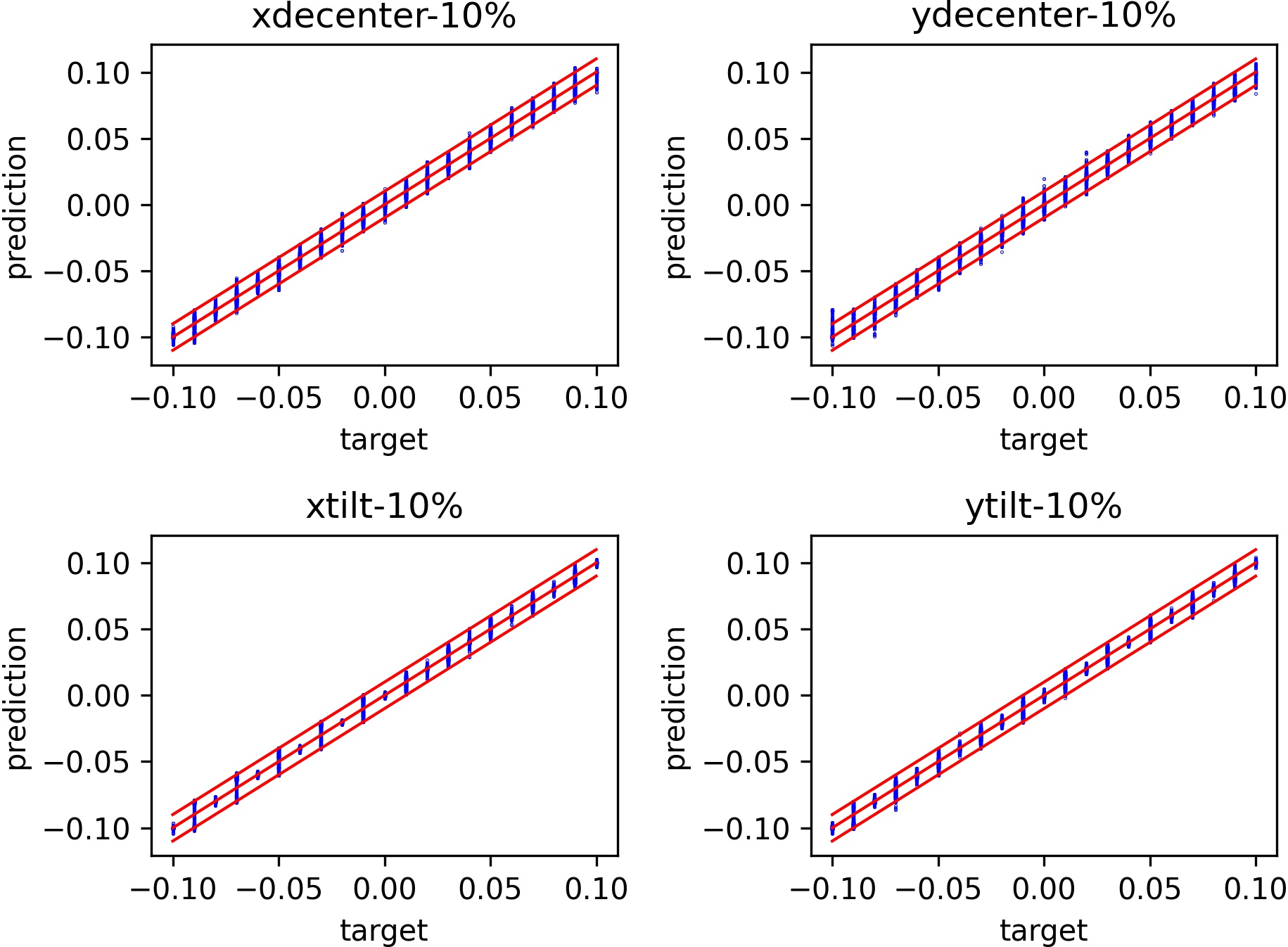}
\caption{Misalignment states prediction results for the R-C telescope with reduced sampling size are presented in this figure. In this figure, we have plotted both the mean value and a $\pm 10\%$ threshold for the misalignment states.}
\label{fig:LPallspaceRC_final0.01movesresult}
\end{figure}

As shown in this figure, our method could effectively estimate misalignment states from PSF maps (error smaller than $10^{-4}$ mm for decenter and $10^{-4}$ degrees for tilt). To provide more specific details, our method achieves a mean value close to $2.859 \times 10^{-4}$ mm for x-decenter, while for x-tilt, the mean value is approximately $3.352\times 10^{-4}$ degrees. In terms of y-decenter, our approach yields a mean value approximating $2.233\times 10^{-4}$ mm, and for y-tilt, the mean value approximates $2.393\times 10^{-4}$ degrees. To delve further into the performance of our neural network in predicting misalignment states not present in the training set, we've reduced the size of the sampling step by a factor of 2 (halving the minimal sampling step). Subsequently, we've employed the network to predict misalignment states from randomly selected sequences. The results are shown in Figure~\ref{fig:LPallspaceRC_final0.01movesresult}. Remarkably, despite the absence of these particular misalignment states in the training set, our method still manages to provide effective results, with errors remaining below $10\%$ of the ground truth values, which is almost the same to errors predicted from the training data. Results obtained in this section support the effectiveness of our method, which uses sequence to decouple aligned misalignment states. Comparing with the results obtained by our group previously \cite{jia2020point}, the new method could simultaneously estimate 4 different misalignment states with error 2 order smaller than the previous method, which only uses a PSF map from one measurement to predicate misalignment states.\\

\subsubsection{Misalignment of the Secondary Mirror in a RC Telescope with five misalignment dimensions}
\begin{figure}[htbp]
\centering\includegraphics[width=6cm]{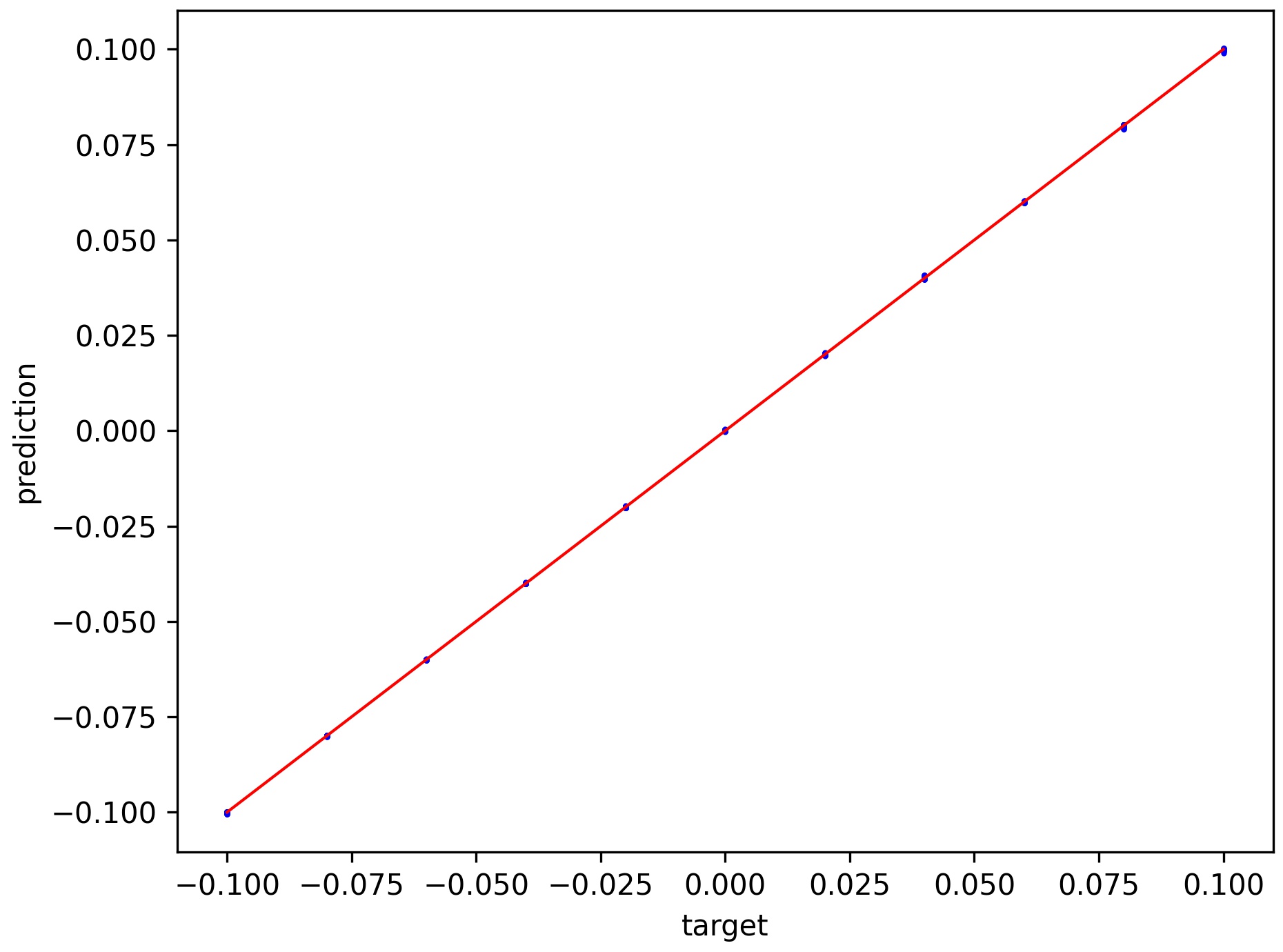}
\caption{Prediction result of decenter on z axis for the secondary mirror of R-C telescope.}
\label{fig:RC5misresult}
\end{figure}

To accommodate more misalignment states, we introduce z-decenter into our set of misalignment dimensions. However, this will lead to a dimension explosion issue, resulting in around 160,000 misalignment states in our misalignment space. We have two options to estimate the z-decenter. The first approach involves generating more data for state prediction. Given the increased number of dimensions, we need to extend the length of the sequence to effectively decouple different states. The other approach is to initially estimate the z-decenter and then continue to estimate the other four misalignment states. We have selected for the second method, which allows us to achieve a relatively high accuracy in our misalignment prediction. For predicting the z-decenter of the secondary mirror in the RC system, we attained a mean squared error (MSE) of less than 1e-04 in 99\% of the sequences, while other dimensions could achieve the same level of accuracy. The prediction results for z-decenter are illustrated in Figure~\ref{fig:RC5misresult}. The results show a mean value of 2.94e-05 and a variance value of 8.19e-10 in millimeters. 

\subsubsection{Misalignment of the Secondary Mirror in a RC Telescope with realistic conditions}
We account for the impact of actual observation conditions in this part, which encompass factors introduced by atmospheric turbulence, dome seeing, atmospheric dispersion, and sub-pixel shifts. To begin, we assess the scenario with the influence of atmospheric turbulence and dome seeing. Atmospheric turbulence and the dome seeing influence fine details of PSFs, making the prediction of certain misalignment states more challenging. In theory, our method remains capable of predicting misalignment states, as it utilizes the complete structure of PSFs for state estimation. In this study, we operate under the assumption that the PSFs introduced by the atmosphere turbulence could be viewed as long-exposure PSFs, which adhere to the Moffat model defined in Equation~\ref{eq2}:
\begin{equation}
\label{eq2}
   \mathrm{PSF}(r)=\frac{\beta-1}{\pi\alpha^2}\left[1+\left(\frac{r}{\alpha}\right)^2\right]^{-\beta},
\end{equation}
where $\alpha$ and $\beta$ are free parameters to describe the PSF. We set $\beta$ to 4.765 for atmospheric turbulence-induced PSFs \cite{trujillo2001effects}. Then, we calculate $\alpha$ with Equation~\ref{eq3}:
\begin{equation}
\label{eq3}
   \alpha = \frac{FWHM}{2\sqrt{{2^{\frac{1}{\beta}} -1}}},
\end{equation}
where $FWHM$ denotes the Full Width at Half Maximum of PSFs. In this paper, $FWHM$ is treated as a random variable, with a mean value of 1.2 arcsec. Meanwhile, the dome seeing is simulated with a thin layer of atmospheric turbulence phase screen with long exposures \citep{jia2015simulation}. To generate realistic PSFs for misalignment estimation, we convolve the PSFs induced by atmospheric turbulence and the PSFs induced by the dome seeing with the optical PSFs.\\

\begin{figure}[htbp]
\centering\includegraphics[width=9cm]{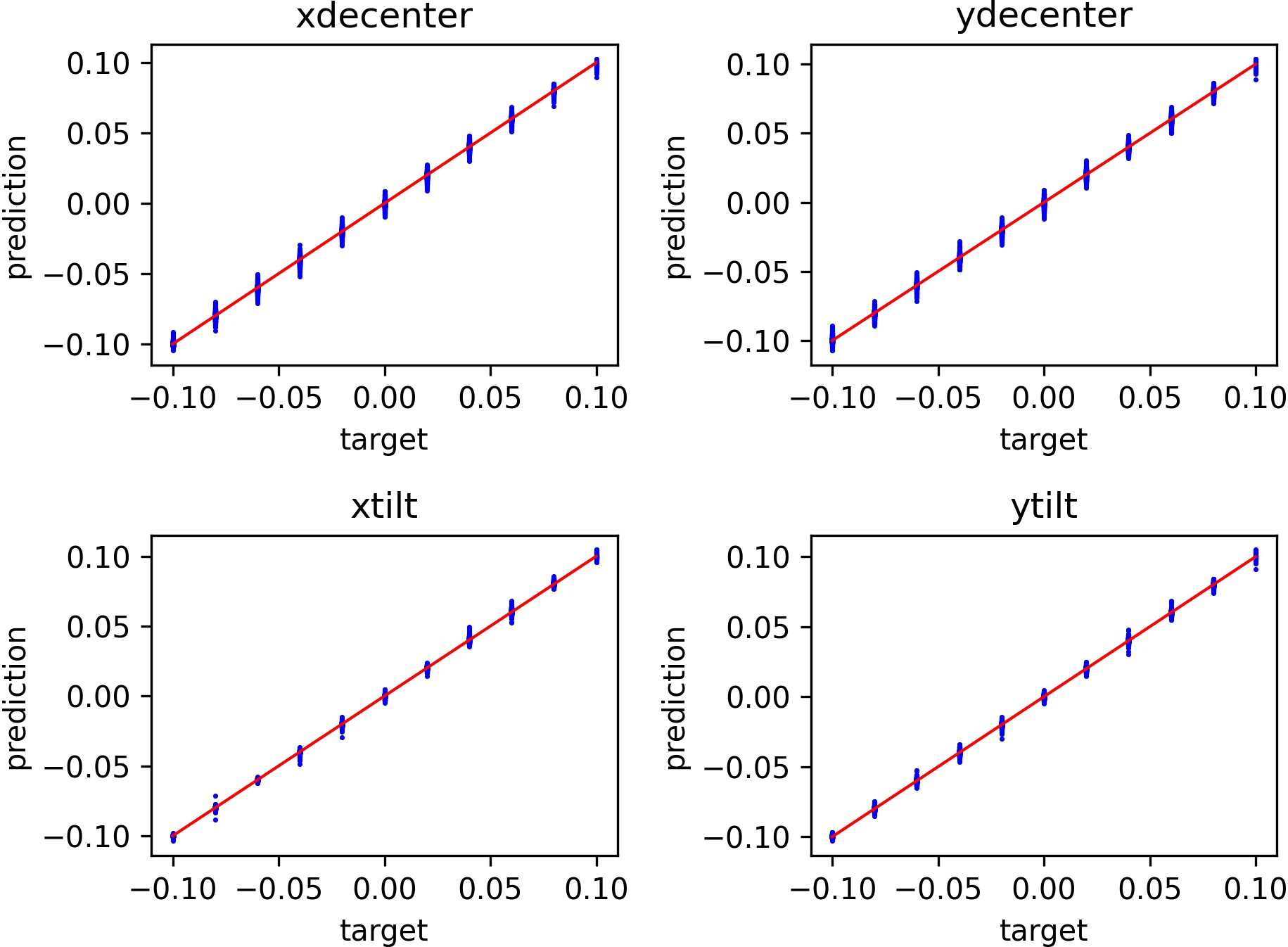}
\caption{Results obtained by our neural network, when the atmospheric turbulence and dome seeing are considered.}
\label{fig:ATPRC-result}
\end{figure}

Because these PSFs are just blurred by the atmospheric turbulence and dome seeing and the main structure do not change significantly, we employ a transfer learning strategy to fine-tune the neural network \cite{zhuang2020comprehensive}. During the transfer learning procedure, we train the neural network with 1000 epochs. The results after transfer learning are illustrated in Figure~\ref{fig:ATPRC-result}. In the test results, some sequences (0.4\%) clearly deviate from the ground truth value due to the influence of atmospheric turbulence and dome seeing. These abnormal points have been excluded from the results. Compared with the tilt and decenter prediction results with the same noise level \cite{jia2020point}, although without excluding the outliers, our network achieves mean value of 2.46e-03, 2.74e-03, 1.05e-03, 1.32e-03 and variance of 1.48e-05, 1.75e-05, 5.18e-06, 6.81e-06 on x-decenter, y-decenter, x-tilt, y-tilt, respectively. The accuracy is improved by an order of magnitude. The overall findings indicate that our method consistently yields satisfactory results, whether in the presence or absence of atmospheric turbulence.\\

\begin{figure}[htbp]
\centering\includegraphics[width=9cm]{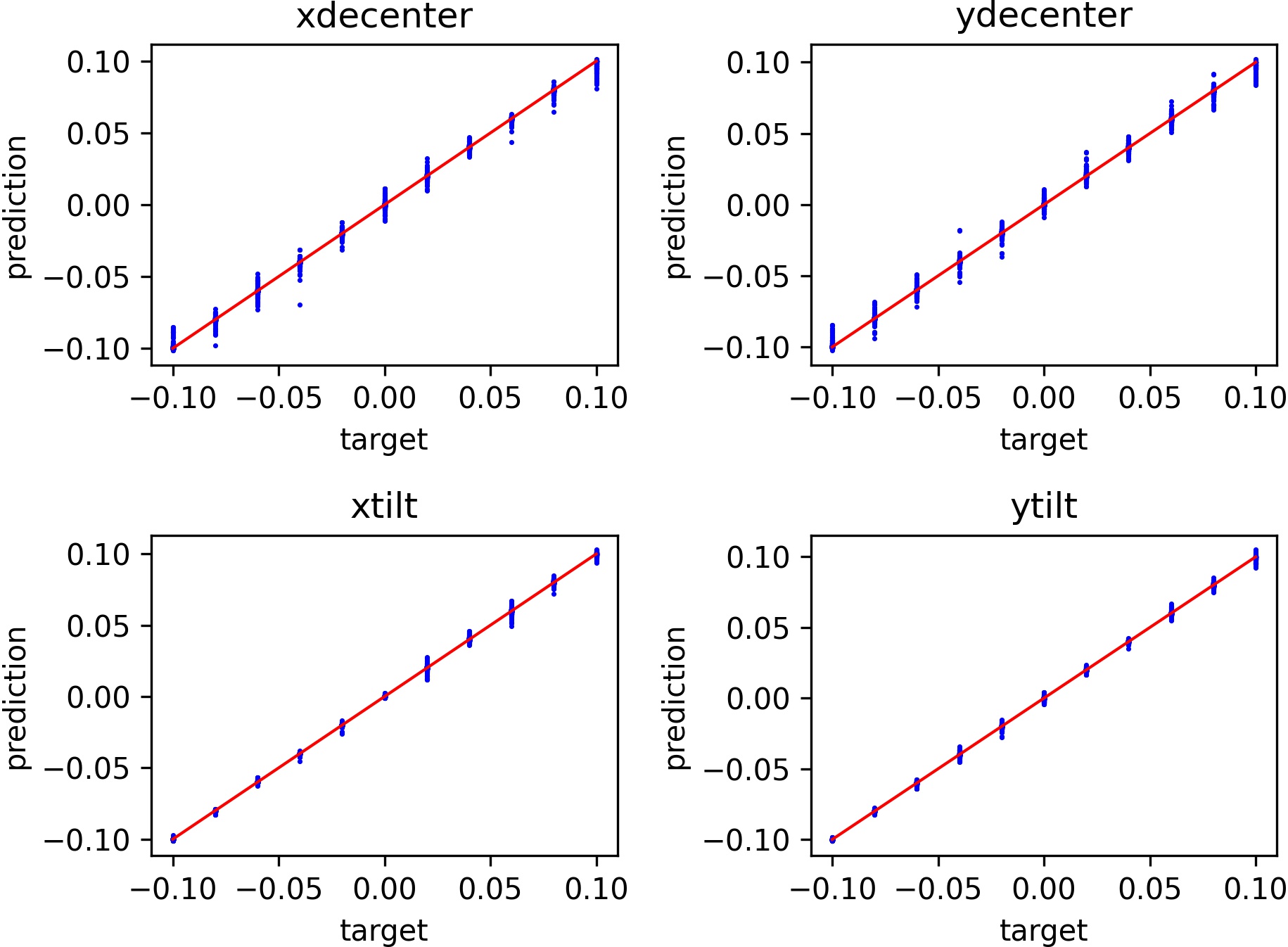}
\caption{Misalignment estimation results of 4 misalignment in RC system after introducing random sub-pixel shift within 0.5 pixel.}
\label{fig:subpixelRC result}
\end{figure}

Secondly, since we could not obtain centered star images, subpixel shift will introduce potential error for misalignment estimation. Therefore, we consider PSFs with random sub-pixel shift with range of 0.5 pixels. We add random sub-pixel shift in the training data and test data. Then we train the neural network from scratch. After training, the results are shown in figure~\ref{fig:subpixelRC result}. It can be seen that sub-pixel shift will have a greater impact on prediction of decenter. The mean value of decenter along x and y are 5.7e-04 mm and 6.7e-04 mm, with variance of 2.1e-06 mm and 2.35e-06 mm. While the mean value of tilt along x and y are 3.1e-04 deg and 3.2e-04 deg, with variance of 3.74e-07 deg and 3.69e-07 deg.\\

\begin{figure}[htbp]
\centering\includegraphics[width=9cm]{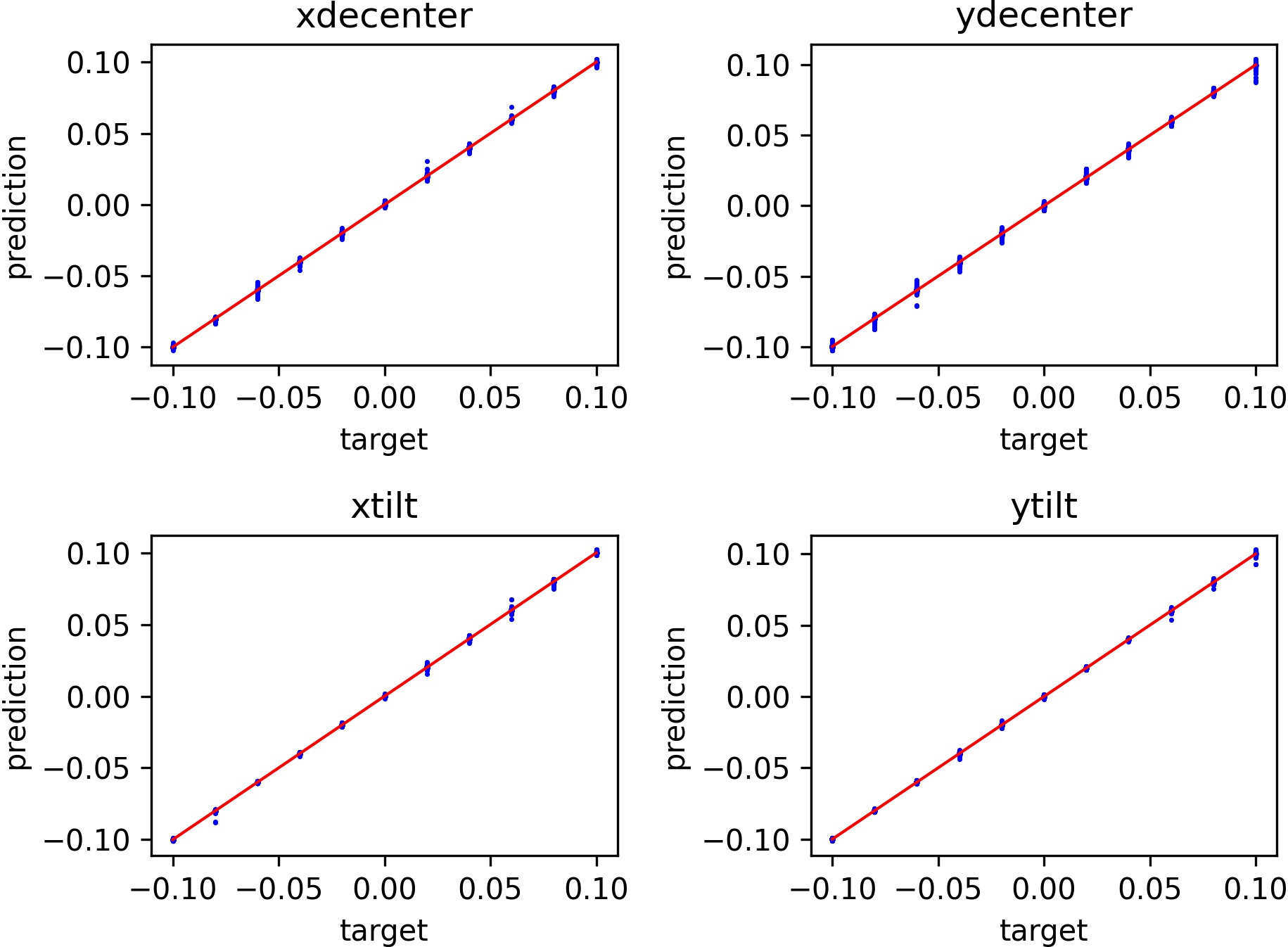}
\caption{Misalignment estimation results of 4 misalignment in RC system after introducing atmospheric dispersion.}
\label{fig:atmospheric dispersion result}
\end{figure}

Thirdly,we consider the impact of the atmosphere dispersion to estimation results. Since the atmosphere dispersion could be corrected by the  atmospheric dispersion corrector \cite{atmospheric_dispersion_corrector}. We consider the residual atmosphere dispersion here. The residual atmosphere dispersion is 0.4 arcsec. In this case, we add Gauss blur with FWHM of 0.4 to simulated PSFs. Then we also use the transfer learning to fine-tune the neural network. The results are show in Figure~\ref{fig:atmospheric dispersion result}. The mean value of decenter along x and y are 3.12e-04 mm and 4.12e-04 mm, with variance of 1.98e-06 mm and 4.13e-06 mm. While the mean value of tilt along x and y are 2.34e-04 deg and 2.24e-04 deg, with variance of 1.08e-07 deg and 8.18e-07 deg. As shown in this example, the atmosphere dispersion will not introduce significant effects to final results.\\

Finally, we consider the vignetting and obstruction problem in final images. We generate PSFs with diffraction calculation and calculate the vignetting coefficients for the whole filed of view. We further assume photons from the target satisfy the Poisson distribution and the back ground satisfies the Gaussian distribution. Since we could normally select targets with enough brightness, we generate images with signal to noise ratio larger than 50 to test the results. We have found that there are no obvious difference between predicted results.\\

\subsection{Scenario II: Misalignment of the Primary Mirror in a PFC Telescope}
\label{Per2}
In the second scenario, we test the performance of our method in estimating misalignment states, focusing on the primary mirror of a PFC telescope. The PFC telescope, known for its wide field of view, is frequently used in sky survey projects. Detailed parameters for the PFC telescope can be found in Table~\ref{tab:PFC parameters}. Estimating the misalignment state of the primary mirror for the PFC telescope poses a more significant challenge compared to estimating the misalignment state of the secondary mirror in the RC telescope. This heightened difficulty is attributed to the more complex optical system of the PFC telescope, which results in complex PSFs when misalignment occurs. In this scenario, we consider four independent misalignment dimensions for the primary mirror: x-decenter, y-decenter, x-tilt, and y-tilt. The decenter range is 0.1 mm with a minimum step size of 0.02 mm, while the tilt range is 0.01 degrees with a minimum step size of 0.002 degrees.\\

\begin{table}[htbp]
\centering
\caption{\bf \centering Parameters of PFC telescope in mm}
\begin{tabular}{ccccc}
\hline
Surface&Radius&Thickness&Semi-Diameter&Glass \\
\hline
    1&-4282.468&-1297.195&500&$MIRROR$\\
    2&-357.903&-50&250.476&$F\underline{~}SILICA$\\
    3&-376.948&-413.099&237.761\\
    4&11704.533&-30&159.266&$F\underline{~}SILICA$\\
    5&-270.946&-74.631&147.579\\
    6&-1384.536&-22&149.341&$H-ZLAF76$\\
    7&-586.686&-40&148.466\\
    8&-337.474&-55&163.843&$H-ZK50$\\
    9&1351.297&0&163.668\\
    10&$Infinity$&-23.864&161.796\\
    11&-430.479&-30&152.084&$CAF2$\\
    12&41863.724&-80&151.604\\
    13&$Infinity$&-12&122.351&$F\underline{~}SILICA$\\
    14&$Infinity$&-16&119.461&\\
    15&$Infinity$&-12&113.630&$F\underline{~}SILICA$\\
    16&$Infinity$&-6&110.740\\
    $17(Image)$&$Infinity$&-&108.561\\
\hline
\end{tabular}
  \label{tab:PFC parameters}
\end{table}

\begin{figure}[htbp]
\centering\includegraphics[width=9cm]{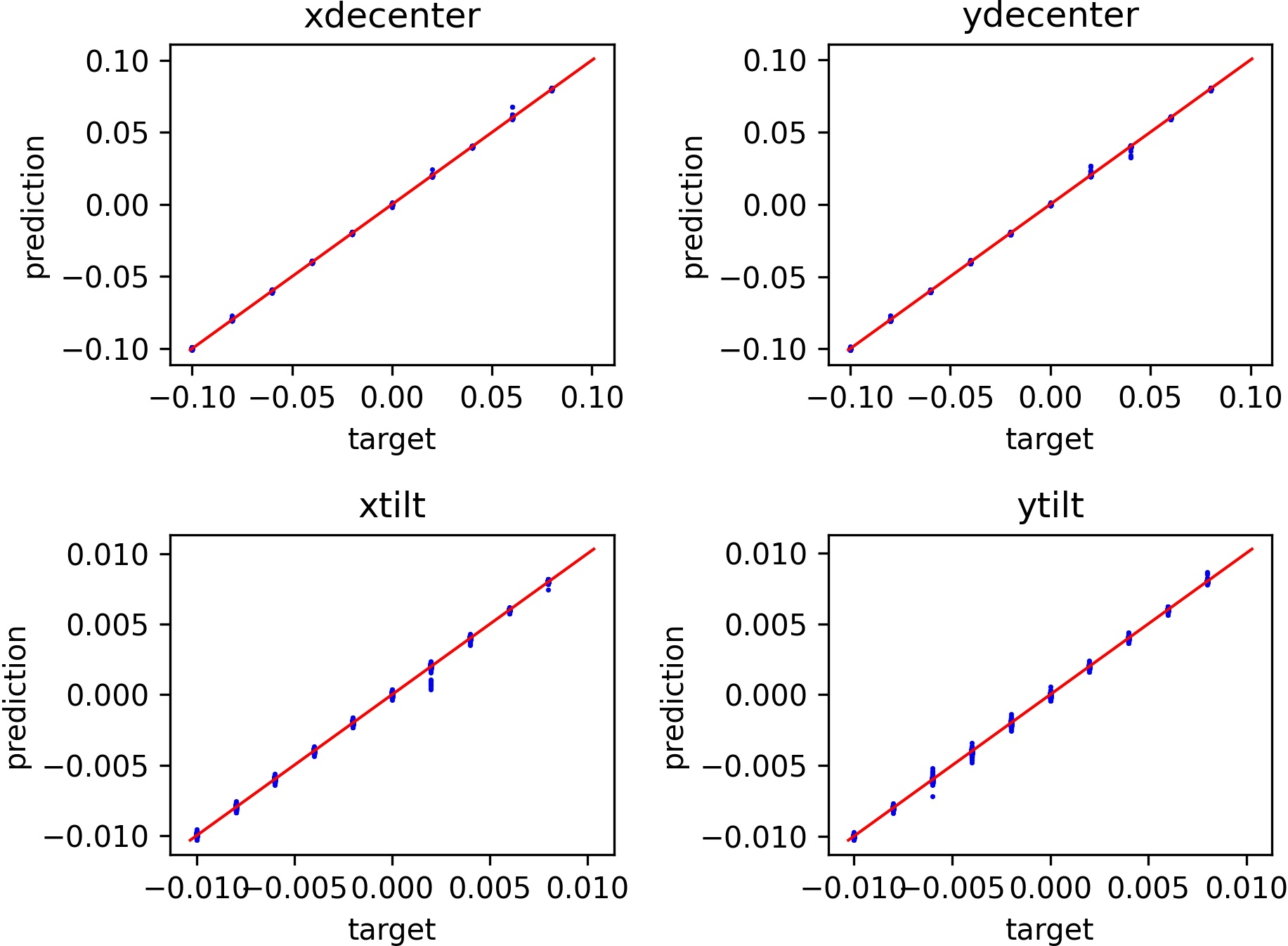}
\caption{The effectiveness of our approach in estimating misalignment states for the primary mirror of the PFC telescope is depicted in this figure.}
\label{fig:allspacePFCresult}
\end{figure}

With the parameters defined above, we utilize the digital twin to generate a state graph to store various misalignment states. Initial training dataset involve 5000 sequences (around 9000 misalignment states inside) along with corresponding PSF maps. In the process of training, we used 2e-5 as learning rate and Adam as optimizer. MSE loss converged within 200 epochs and gradually decrease to reach higher accuracy. The neural network has been trained by 4000 epochs and cost 550 hours using sequence with length of 5 (5 continuous misalignment states) derived from this state graph. During the training stage, due to different scales of decenter and tilt, we scale the target values and predicted values of tilt by a factor of 10. After the training stage, we randomly select 1000 sequences from the state graph for testing, and the results obtained through our method are displayed in Figure~\ref{fig:allspacePFCresult}. The mean errors for x-decenter, y-decenter, x-tilt, and y-tilt are 7.72e-05 mm, 8.69e-05 mm, 2.45e-05 degrees, and 2.07e-05 degrees, respectively. In this figure, the blue dots in the decenter figures are closer to the red line compared to those in the tilt figures. This discrepancy is due to the larger step size of decenter, which makes it easier to be predicted. As illustrated, our method effectively estimates misalignment states of the primary mirror.\\

\begin{figure}[htbp]
\centering\includegraphics[width=9cm]{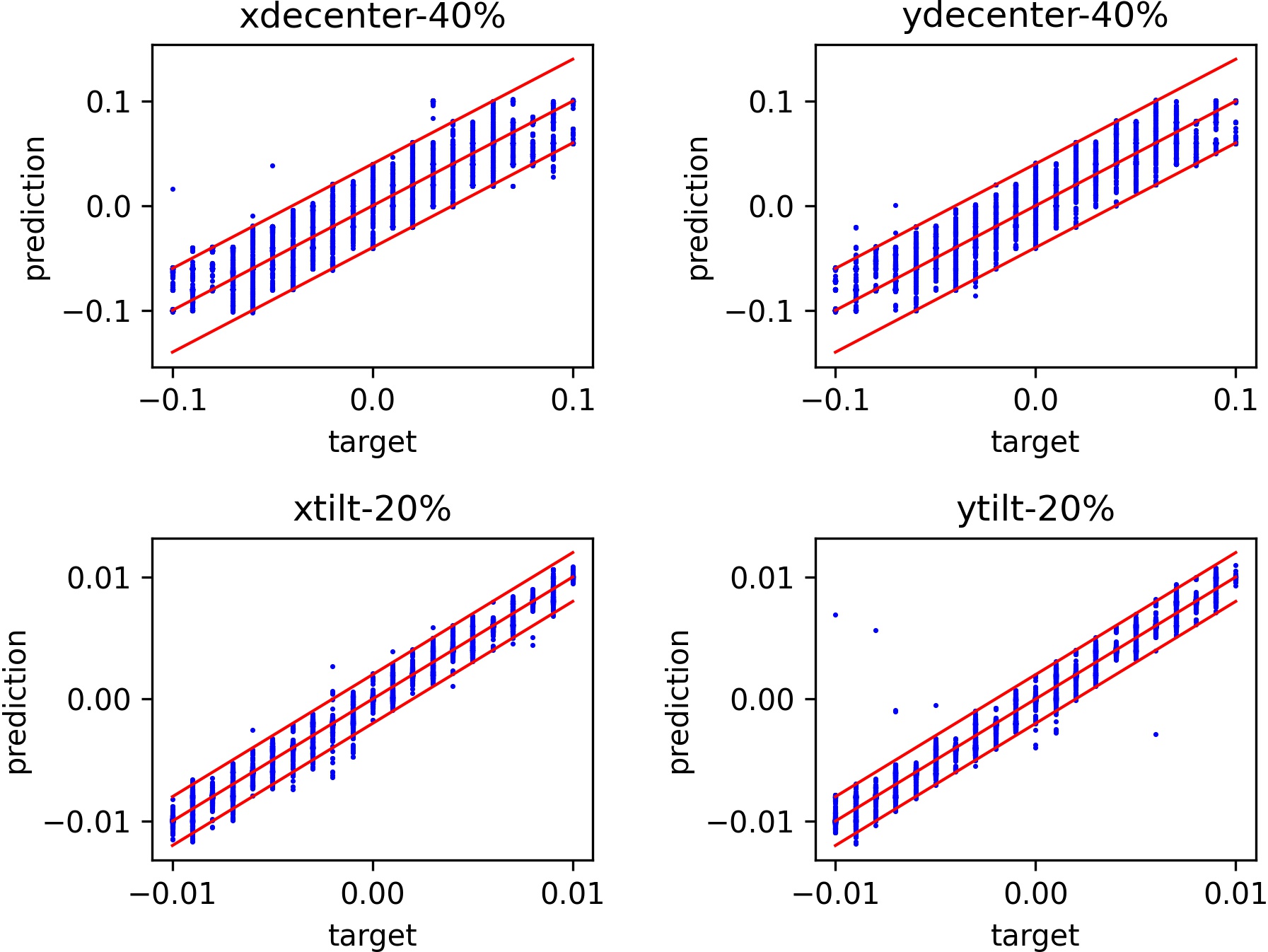}
\caption{The prediction results obtained by our method, when the sampling step size is reduced by a factor of 2. The top two figures show estimation results of decenter along x and y with $\pm 40 \%$ error and tilt along x and y with $\pm 20 \%$ error.}
\label{fig:LPPFC_0.01result}
\end{figure}

We have further reduced the sampling step size by a factor of 2 and randomly selected several sequences to assess the performance of our method. Importantly, no transfer-learning is conducted in this phase, as we aim to investigate whether our method could still yield effective results without fine-tuning. The results are presented in Figure~\ref{fig:LPPFC_0.01result}. As depicted in the figure, even after reducing the step size by half, our method continues to produce effective results, albeit with increased error rates. In particular, the decenter error remains within $\pm 40\%$, while the tilt error falls within $\pm 20\%$. Based on sensitivity analysis, it is worth noting that the contribution of tilt to the RMS of PSF is three orders of magnitude greater than that of decenter when the unit of decenter is in mm and the unit of tilt is in degrees. Therefore, the accuracy of predicted tilt tends to be slightly higher after reducing the step size. Given the inherently complex optical structure of the PFC system, the misalignment error is not as low as that achieved in the RC telescope.\\

\begin{figure}[htbp]
\centering\includegraphics[width=6cm]{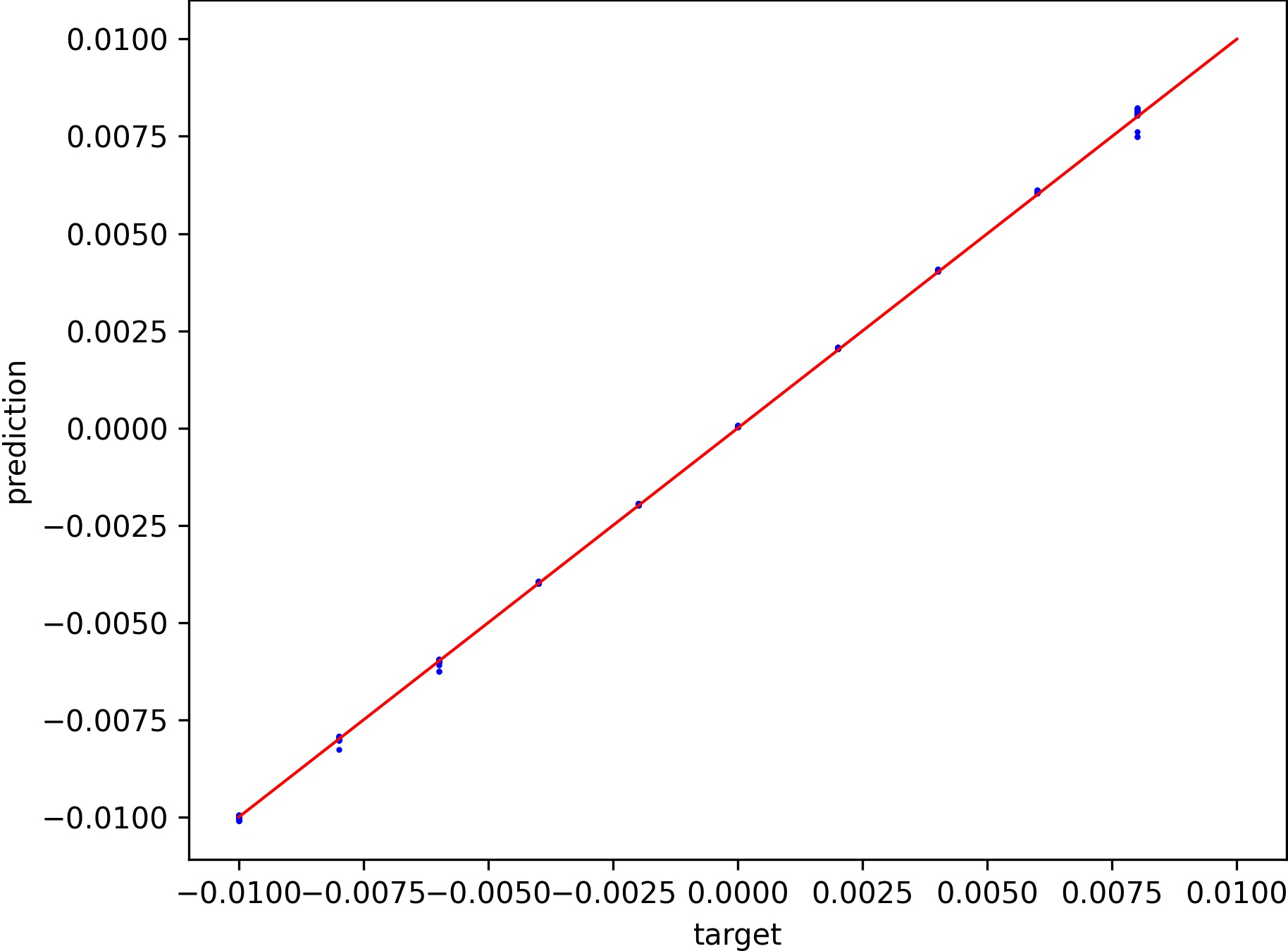}
\caption{Prediction result of decenter on z axis for the primary mirror of PFC telescope.}
\label{fig:PFC5misresult}
\end{figure}

At last, we investigate the case, which contains misalignment of x tilt, x decenter, y tilt, y decenter and z decenter. We use the same strategy as mentioned in Section~\ref{Per1}. Then we could achieve that 97\% sequences out of 20000 sequences have relatively good result (MSE less than 1e-07) as shown in Figure~\ref{fig:PFC5misresult}. The result could reach a mean value of 3.33e-05 and variance of 6.61e-10 in mm.

\subsection{Scenario III: Misalignment of the Primary Mirror and the Secondary Mirror in an RC Telescope}\label{2mirroronlytilt}
Finally, we evaluate the performance of our method using an RC telescope with misalignment in both the primary and secondary mirrors. Given the multitude of potential misalignment states, particularly when considering two optical elements with six misalignment dimensions (tilt and decenter along the x, y, and z axes), we have narrowed our focus to the tilt in the x and y directions for both the primary and secondary mirrors.\\

\begin{figure}[htbp]
\centering\includegraphics[width=9cm]{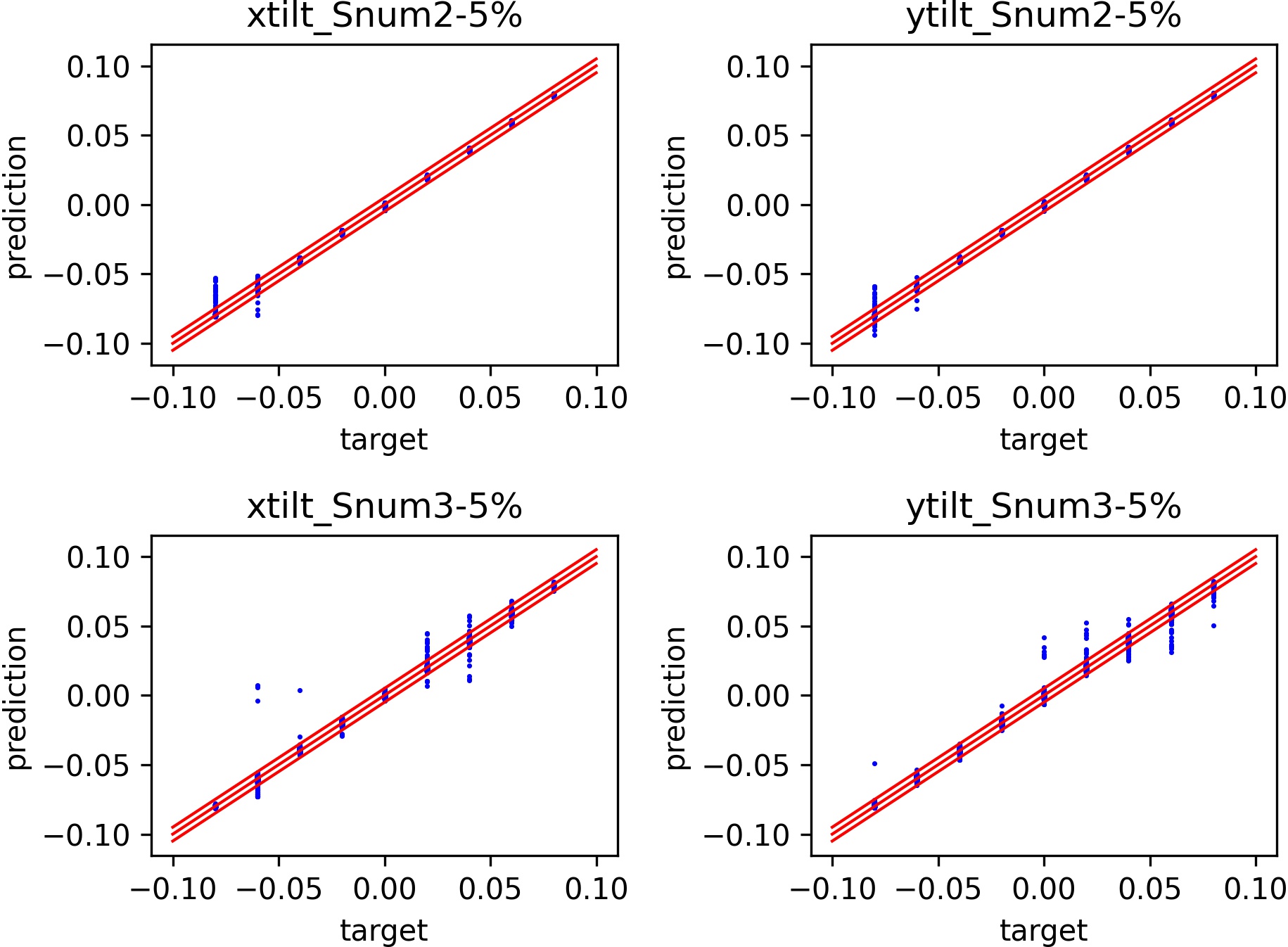}
\caption{Results for estimating the misalignment states of a RC telescope with both the primary (Snum:2) and secondary mirrors (Snum:3) experiencing tilt misalignment are presented. We've illustrated the mean values and marked the $\pm 5\%$ error margins with red lines.}
\label{fig:len5_2mirroronlytilt_result}
\end{figure}

Initially, we apply the same approach as described earlier, acquiring a batch of misalignment sequences with a length of 5 (around 3500 misalignment states involved inside) to train the neural network, which cost 1500 hours (8000 epochs) to train. During the training stage, 2e-5 of learning rate and Adam optimizer are used. For the dataset with sequence length of 5, MSE loss drops sharply within 60 epochs and tends to converge, then MSE continues to decrease until we get optimal model. However, for the two-mirror tilt dataset with a sequence length of 10, the MSE drops sharply within 40 epochs, and then the loss curve shows a downward straight line with a slight slope. Subsequently, we employ the trained neural network to estimate misalignment states. The results are shown in Figure~\ref{fig:len5_2mirroronlytilt_result}. Despite the substantial reduction in MSE to a very low level (6.89e-08) during the training stage, the estimated results still exhibit noticeable deviations from the ground truth values.\\

\begin{figure}[htbp]
\centering\includegraphics[width=9cm]{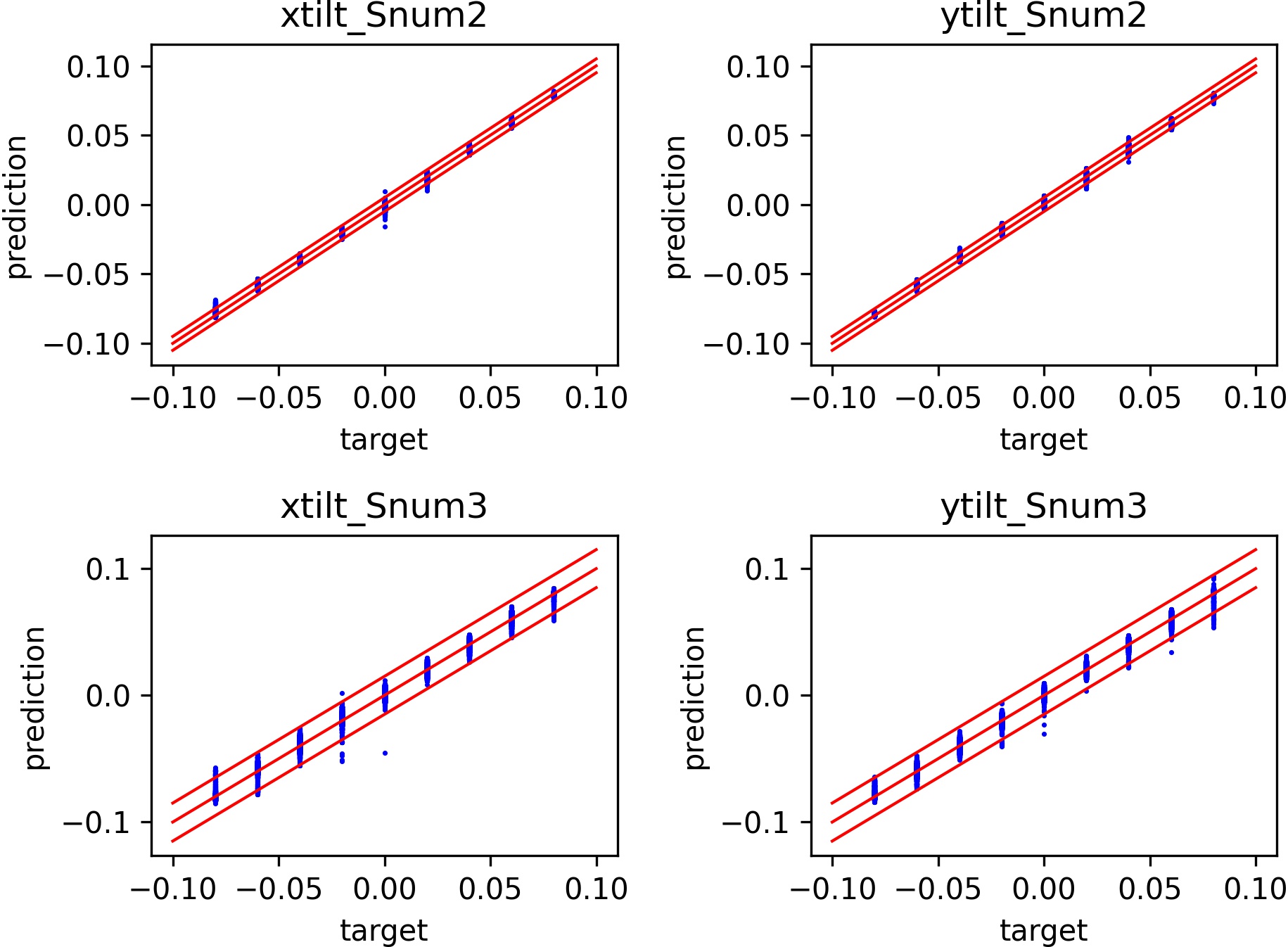}
\caption{Results for estimating the misalignment states of a RC telescope with both the primary (Snum:2) and secondary mirrors (Snum:3) experiencing tilt misalignment are presented. We've illustrated the mean values and marked the $\pm 5\%$ error margins with red lines in top figures and $\pm 15 \%$ in bottom figures.}
\label{fig:len1_2mirroronlytilt_result}
\end{figure}

The deviation may arise from the coupling of the misalignment in different directions. To mitigate this issue, we opt to increase the length of the sequence, effectively addressing the coupling problem. Subsequently, we extend the sequence length and select a batch of sequences with a length of 10 for neural network training which introduce around 1800 more misalignment states. The neural network has been trained by 2000 epochs and cost 350 hours. After training, we randomly select several sequences to estimate misalignment states, and the results are visualized in Figure~\ref{fig:len1_2mirroronlytilt_result}. As depicted in this figure, extending the sequence length brings the estimation results closer to the ground-truth values, albeit at the expense of increased computational resources. Furthermore, the model performs notably well for the primary mirror case (num-2), with errors primarily concentrated within a $5\%$ margin of the threshold. In contrast, for the secondary mirror (num-3), errors are generally within a $15\%$ margin of the threshold. This suggests that misalignment of the primary mirror will have more impacts to full-field-of-view PSFs. In conclusion, even with more optical elements in a telescope with misalignment, our method could still return effective results through increasing the length of misalignment state sequence.\\

\section{Conclusion}\label{conclusion}
In recent years, sky survey telescopes have become increasingly crucial for astronomical observations, especially in time domain astronomy, where multiple telescopes function automatically as a telescope array to capture celestial object images \cite{xu2020real,danieli2020dragonfly,liu2021sitian,law2022low,ofek2023large}. Maintaining and controlling these telescopes present a challenge due to limited human resources and tight time constraints \cite{jia2023simulation, jia2023observation}. Recognizing misalignment states becomes the foremost and critical step in telescope maintenance. Since these telescopes lack wavefront sensors, it becomes imperative to develop an automated method capable of detecting misalignment states solely based on observation data. By doing so, we can efficiently maintain these telescopes with minimal human intervention, ensuring their optimal performance. For larger telescopes, which normally equipped with active optics system and wavefront sensor, the digital twin could use history data to provide an initial guess of the telescope states. Then we can modify the structure of the telescope in a more efficient way.\\

Considering these requirements, we propose an innovative method for detecting misalignment states from observation data. Our approach assumes that misalignment states and their corresponding PSFs (PSF maps in this paper) can be accurately simulated using the digital twin. By leveraging a sequence of PSF maps instead of a single PSF map, we can enhance the accuracy of misalignment predictions with neural networks. With this concept in mind, we outline our method as follows. Firstly, we utilise the digital twin to generate a substantial amount of data, comprising misalignment states and their corresponding PSF maps. This data is stored using a state graph, and we actively construct sequences containing adjacent misalignment states to train a neural network. After training, our method effectively identifies misalignment states in most cases. For certain coupled misalignment states, we utilize outliers in the state graph to describe limitations of our method. We validate our method through three different scenarios and the results demonstrate its effectiveness. \\

Our method actively bridges the digital twin of telescopes with test data from the laboratory. With the combined power of the state graph and the deep neural network, we obtain an effective strategy for estimating misalignment. After training, the state graph, together with the neural network, guides us in acquiring useful test data from the laboratory. These real data can then be used to fine-tune the neural network, making it applicable to real observations. Although the framework has proven effective, there are still areas that can be further improved:
\begin{itemize}
    \item First, to gather a large amount of data from the digital twin, it is essential to develop highly efficient and high-fidelity simulation methods. Current commercial and open-source optical simulation methods are too slow. As a result, our group is currently working on developing physically-inspired neural networks for simulating optical systems.
    \item Second, since we need to adjust optical components in the optical system to obtain extensive test data, it is crucial to employ more efficient sampling methods to obtain data of telescopes with different misalignment states, such as reinforcement learning.
    \item Third, considering the maintenance requirements, we can consider new telescope designs that simplify the data collection and misalignment estimation processes.
    \item Finally, as obtaining precise positions in experiments poses a challenge, we must take these uncertainties into careful consideration in our future endeavors.
\end{itemize}

\section{Acknowledgements}
This work is supported by the National Natural Science Foundation of China (NSFC) with funding numbers 12173027 and 12173062. We acknowledge the science research grants from the China Manned Space Project with NO. CMS-CSST-2021-A01. The authors acknowledge the science research grants from the Square Kilometer Array (SKA) project with NO. 2020SKA0110102. Specially this work is funded by the Major Key Project of PCL.

\section*{Data Availability}
The code can be downloaded from the PaperData Repository powered by China-VO (\url{doi:10.12149/101334}) .\\


\bibliography{main}

\begin{thebibliography}{10}
\newcommand{\enquote}[1]{``#1''}

\bibitem{Veran1997}
J.~P. {Veran}, F.~{Rigaut}, H.~{Maitre}, and D.~{Rouan}, \enquote{{Estimation of the adaptive optics long-exposure point-spread function using control loop data.}} {\protect\JournalTitle{Journal of the Optical Society of America A}} \textbf{14}, 3057--3069 (1997).

\bibitem{Gendron2006}
E.~{Gendron}, Y.~{Cl{\'e}net}, T.~{Fusco}, and G.~{Rousset}, \enquote{{New algorithms for adaptive optics point-spread function reconstruction},} {\protect\JournalTitle{Astronomy and Astrophysics}} \textbf{457}, 359--363 (2006).

\bibitem{Jolissaint2012}
L.~{Jolissaint}, C.~{Neyman}, J.~{Christou}, and P.~{Wizinowich}, \enquote{{Adaptive optics point spread function reconstruction project at W. M. Keck Observatory: first results with faint natural guide stars},} in \emph{Adaptive Optics Systems III,}  vol. 8447 of \emph{Society of Photo-Optical Instrumentation Engineers (SPIE) Conference Series} B.~L. {Ellerbroek}, E.~{Marchetti}, and J.-P. {V{\'e}ran}, eds. (2012), p. 844728.

\bibitem{Martin2016}
O.~A. {Martin}, C.~M. {Correia}, E.~{Gendron}, \emph{et~al.}, \enquote{{PSF reconstruction validated using on-sky CANARY data in MOAO mode},} in \emph{Adaptive Optics Systems V,}  vol. 9909 of \emph{Society of Photo-Optical Instrumentation Engineers (SPIE) Conference Series} E.~{Marchetti}, L.~M. {Close}, and J.-P. {V{\'e}ran}, eds. (2016), p. 99091Q.

\bibitem{Wagner2018}
R.~{Wagner}, C.~{Hofer}, and R.~{Ramlau}, \enquote{{Point spread function reconstruction for single-conjugate adaptive optics on extremely large telescopes},} {\protect\JournalTitle{Journal of Astronomical Telescopes, Instruments, and Systems}} \textbf{4}, 049003 (2018).

\bibitem{Turri2022}
P.~{Turri}, J.~R. {Lu}, G.~{Witzel}, \emph{et~al.}, \enquote{{AIROPA III: testing simulated and on-sky data},} {\protect\JournalTitle{Journal of Astronomical Telescopes, Instruments, and Systems}} \textbf{8}, 039002 (2022).

\bibitem{Moffat1969}
A.~F.~J. {Moffat}, \enquote{{A Theoretical Investigation of Focal Stellar Images in the Photographic Emulsion and Application to Photographic Photometry},} {\protect\JournalTitle{Astronomy and Astrophysics}} \textbf{3}, 455 (1969).

\bibitem{Kormendy1973}
J.~{Kormendy}, \enquote{{Calibration of direct photographs using brightness profiles of field stars.}} {\protect\JournalTitle{Astronomical Journal}} \textbf{78}, 255--262 (1973).

\bibitem{Stetson1992}
P.~B. {Stetson}, \enquote{{More Experiments with DAOPHOT II and WF/PC Images},} in \emph{Astronomical Data Analysis Software and Systems I,}  vol.~25 of \emph{Astronomical Society of the Pacific Conference Series} D.~M. {Worrall}, C.~{Biemesderfer}, and J.~{Barnes}, eds. (1992), p. 297.

\bibitem{Jee2011}
M.~J. {Jee} and J.~A. {Tyson}, \enquote{{Toward Precision LSST Weak-Lensing Measurement. I. Impacts of Atmospheric Turbulence and Optical Aberration},} {\protect\JournalTitle{Publications of the Astronomical Society of the Pacific}} \textbf{123}, 596 (2011).

\bibitem{ma2008diagnosing}
Z.~Ma, G.~Bernstein, A.~Weinstein, and M.~Sholl, \enquote{Diagnosing space telescope misalignment and jitter using stellar images,} {\protect\JournalTitle{Publications of the Astronomical Society of the Pacific}} \textbf{120}, 1307 (2008).

\bibitem{schechter2011generic}
P.~L. Schechter and R.~S. Levinson, \enquote{Generic misalignment aberration patterns in wide-field telescopes,} {\protect\JournalTitle{Publications of the Astronomical Society of the Pacific}} \textbf{123}, 812 (2011).

\bibitem{li2015alignment}
Z.~Li, X.~Yuan, and X.~Cui, \enquote{Alignment metrology for the antarctica kunlun dark universe survey telescope,} {\protect\JournalTitle{Monthly Notices of the Royal Astronomical Society}} \textbf{449}, 425--430 (2015).

\bibitem{li2020active}
Z.~Li, X.~Yuan, K.~Zhang, and B.~Li, \enquote{Active alignment metrology for multi-channel photometric survey telescope,} in \emph{Advances in Optical Astronomical Instrumentation 2019,}  vol. 11203 (SPIE, 2020), pp. 13--14.

\bibitem{an2021alignment}
Q.~An, X.~Wu, X.~Lin, \emph{et~al.}, \enquote{Alignment of decam-like large survey telescope for real-time active optics and error analysis,} {\protect\JournalTitle{Optics Communications}} \textbf{484}, 126685 (2021).

\bibitem{anqichang_internal_motion_metrology_system}
Q.~An, H.~Zhang, X.~Wu, \emph{et~al.}, \enquote{Photonics large-survey telescope internal motion metrology system,} {\protect\JournalTitle{Photonics}} \textbf{10}, 595 (2023).

\bibitem{bai2021active}
X.~Bai, G.~Ju, B.~Xu, \emph{et~al.}, \enquote{Active alignment of space astronomical telescopes by matching arbitrary multi-field stellar image features,} {\protect\JournalTitle{Optics Express}} \textbf{29}, 24446--24465 (2021).

\bibitem{wu2022machine}
Z.~Wu, Y.~Zhang, R.~Tang, \emph{et~al.}, \enquote{Machine learning for improving stellar image-based alignment in wide-field telescopes,} {\protect\JournalTitle{Research in Astronomy and Astrophysics}} \textbf{22}, 015008 (2022).

\bibitem{zhang2022method}
X.~Zhang, P.~Jia, and W.~Wang, \enquote{A method to build digital twin of atmospheric turbulence phase screens with comprehensible deep neural networks,} in \emph{Adaptive Optics Systems VIII,}  vol. 12185 (SPIE, 2022), pp. 1065--1070.

\bibitem{jia2020point}
P.~Jia, X.~Li, Z.~Li, \emph{et~al.}, \enquote{Point spread function modelling for wide-field small-aperture telescopes with a denoising autoencoder,} {\protect\JournalTitle{Monthly Notices of the Royal Astronomical Society}} \textbf{493}, 651--660 (2020).

\bibitem{jia2021point}
P.~Jia, X.~Wu, Z.~Li, \emph{et~al.}, \enquote{Point spread function estimation for wide field small aperture telescopes with deep neural networks and calibration data,} {\protect\JournalTitle{Monthly Notices of the Royal Astronomical Society}} \textbf{505}, 4717--4725 (2021).

\bibitem{bellman2015applied}
R.~E. Bellman and S.~E. Dreyfus, \emph{Applied dynamic programming}, vol. 2050 (Princeton university press, 2015).

\bibitem{oteo2013new}
E.~Oteo and J.~Arasa, \enquote{New strategy for misalignment calculation in optical systems using artificial neural networks,} {\protect\JournalTitle{Optical Engineering}} \textbf{52}, 074105 (2013).

\bibitem{liu2020misalignment}
Z.~Liu, Q.~Peng, Y.~Xu, \emph{et~al.}, \enquote{Misalignment calculation on off-axis telescope system via fully connected neural network,} {\protect\JournalTitle{IEEE Photonics Journal}} \textbf{12}, 1--12 (2020).

\bibitem{angel2001lsst}
J.~Angel, C.~Claver, R.~Sarlot, \emph{et~al.}, \enquote{Lsst optical design,} in \emph{American Astronomical Society Meeting Abstracts,}  vol. 199 (2001), pp. 101--07.

\bibitem{hodapp2004design}
K.~Hodapp, N.~Kaiser, H.~Aussel, \emph{et~al.}, \enquote{Design of the pan-starrs telescopes,} {\protect\JournalTitle{Astronomische Nachrichten: Astronomical Notes}} \textbf{325}, 636--642 (2004).

\bibitem{ackermann2007large}
M.~R. Ackermann and J.~T. McGraw, \enquote{Large-aperture, three-mirror telescopes for near-earth space surveillance: A look from the outside in,} in \emph{Proceedings of the 2007 AMOS Technical Conference E,}  vol.~6 (Citeseer, 2007).

\bibitem{yuan2012optical}
X.~Yuan and D.-q. Su, \enquote{Optical system of the three antarctic survey telescopes,} {\protect\JournalTitle{Monthly Notices of the Royal Astronomical Society}} \textbf{424}, 23--30 (2012).

\bibitem{grieves2017digital}
M.~Grieves and J.~Vickers, \enquote{Digital twin: Mitigating unpredictable, undesirable emergent behavior in complex systems,} in \emph{Transdisciplinary perspectives on complex systems,}  (Springer, 2017), pp. 85--113.

\bibitem{zhuang2018digital}
C.~Zhuang, J.~Liu, and H.~Xiong, \enquote{Digital twin-based smart production management and control framework for the complex product assembly shop-floor,} {\protect\JournalTitle{The international journal of advanced manufacturing technology}} \textbf{96}, 1149--1163 (2018).

\bibitem{Li2020fast}
Q.-W. {Li}, P.~{Jiang}, and H.~{Li}, \enquote{{Prognostics and health management of FAST cable-net structure based on digital twin technology},} {\protect\JournalTitle{Research in Astronomy and Astrophysics}} \textbf{20}, 067 (2020).

\bibitem{bednarz2020digital}
T.~Bednarz, D.~Branchaud, F.~Wang, \emph{et~al.}, \enquote{Digital twin of the australian square kilometre array (askap),} in \emph{SIGGRAPH Asia 2020 Posters,}  (2020), pp. 1--2.

\bibitem{jia2022digital}
P.~Jia, W.~Wang, R.~Ning, and X.~Xue, \enquote{Digital twin of atmospheric turbulence phase screens based on deep neural networks,} {\protect\JournalTitle{Optics Express}} \textbf{30}, 21362--21376 (2022).

\bibitem{wang2012computer}
L.~Wang and B.~Ellerbroek, \enquote{Computer simulations and real-time control of elt ao systems using graphical processing units,} in \emph{Adaptive Optics Systems III,}  vol. 8447 (SPIE, 2012), pp. 780--790.

\bibitem{rigaut2013simulating}
F.~Rigaut and M.~Van~Dam, \enquote{Simulating astronomical adaptive optics systems using yao,} in \emph{Proceedings of the Third AO4ELT Conference,}  (2013), p.~18.

\bibitem{conan2014object}
R.~Conan and C.~Correia, \enquote{Object-oriented matlab adaptive optics toolbox,} in \emph{Adaptive optics systems IV,}  vol. 9148 (SPIE, 2014), pp. 2066--2082.

\bibitem{peterson2015simulation}
J.~Peterson, J.~Jernigan, S.~Kahn, \emph{et~al.}, \enquote{Simulation of astronomical images from optical survey telescopes using a comprehensive photon monte carlo approach,} {\protect\JournalTitle{The Astrophysical Journal Supplement Series}} \textbf{218}, 14 (2015).

\bibitem{perrin2016poppy}
M.~Perrin, J.~Long, E.~Douglas, \emph{et~al.}, \enquote{Poppy: physical optics propagation in python,} {\protect\JournalTitle{Astrophysics Source Code Library}} pp. ascl--1602 (2016).

\bibitem{reeves2016soapy}
A.~Reeves, \enquote{Soapy: an adaptive optics simulation written purely in python for rapid concept development,} in \emph{Adaptive Optics Systems V,}  vol. 9909 (SPIE, 2016), pp. 2173--2183.

\bibitem{por2018high}
E.~H. Por, S.~Y. Haffert, V.~M. Radhakrishnan, \emph{et~al.}, \enquote{High contrast imaging for python (hcipy): an open-source adaptive optics and coronagraph simulator,} in \emph{Adaptive Optics Systems VI,}  vol. 10703 (SPIE, 2018), pp. 1112--1125.

\bibitem{basden2018durham}
A.~G. Basden, N.~Bharmal, D.~Jenkins, \emph{et~al.}, \enquote{The durham adaptive optics simulation platform (dasp): Current status,} {\protect\JournalTitle{SoftwareX}} \textbf{7}, 63--69 (2018).

\bibitem{ferreira2018compass}
F.~Ferreira, D.~Gratadour, A.~Sevin, and N.~Doucet, \enquote{Compass: an efficient gpu-based simulation software for adaptive optics systems,} in \emph{2018 International Conference on High Performance Computing \& Simulation (HPCS),}  (IEEE, 2018), pp. 180--187.

\bibitem{ren2020thermal}
K.~Ren, Y.~Chew, Y.~Zhang, \emph{et~al.}, \enquote{Thermal field prediction for laser scanning paths in laser aided additive manufacturing by physics-based machine learning,} {\protect\JournalTitle{Computer Methods in Applied Mechanics and Engineering}} \textbf{362}, 112734 (2020).

\bibitem{han2021dnn}
S.~Han, H.-S. Choi, J.~Choi, \emph{et~al.}, \enquote{A dnn-based data-driven modeling employing coarse sample data for real-time flexible multibody dynamics simulations,} {\protect\JournalTitle{Computer Methods in Applied Mechanics and Engineering}} \textbf{373}, 113480 (2021).

\bibitem{trujillo2001effects}
I.~Trujillo, J.~Aguerri, J.~Cepa, and C.~Gutierrez, \enquote{The effects of seeing on sersic profiles--ii. the moffat psf,} {\protect\JournalTitle{Monthly Notices of the Royal Astronomical Society}} \textbf{328}, 977--985 (2001).

\bibitem{jia2015simulation}
P.~Jia, D.~Cai, D.~Wang, and A.~Basden, \enquote{Simulation of atmospheric turbulence phase screen for large telescope and optical interferometer,} {\protect\JournalTitle{Monthly Notices of the Royal Astronomical Society}} \textbf{447}, 3467--3474 (2015).

\bibitem{su2012atmospheric}
D.-q. Su, P.~Jia, and G.~Liu, \enquote{Atmospheric dispersion corrector for the large sky area multi-object fibre spectroscopic telescope,} {\protect\JournalTitle{Monthly Notices of the Royal Astronomical Society}} \textbf{419}, 3406--3413 (2012).

\bibitem{mcinnes2018umap}
L.~McInnes, J.~Healy, N.~Saul, and L.~Gro{\ss}berger, \enquote{Umap: Uniform manifold approximation and projection,} {\protect\JournalTitle{Journal of Open Source Software}} \textbf{3}, 861 (2018).

\bibitem{ambatipudi2023comparison}
S.~Ambatipudi and S.~Byna, \enquote{A comparison of hdf5, zarr, and netcdf4 in performing common i/o operations,}  (2023).

\bibitem{jia2018ground}
P.~Jia, A.~Basden, and J.~Osborn, \enquote{Ground-layer adaptive-optics system modelling for the chinese large optical/infrared telescope,} {\protect\JournalTitle{Monthly Notices of the Royal Astronomical Society}} \textbf{479}, 829--843 (2018).

\bibitem{hughes2023coma}
J.~M. Hughes, C.~E. DeForest, and D.~B. Seaton, \enquote{Coma off it: Regularizing variable point-spread functions,} {\protect\JournalTitle{The Astronomical Journal}} \textbf{165}, 204 (2023).

\bibitem{he2016deep}
K.~He, X.~Zhang, S.~Ren, and J.~Sun, \enquote{Deep residual learning for image recognition,} in \emph{Proceedings of the IEEE conference on computer vision and pattern recognition,}  (2016), pp. 770--778.

\bibitem{ELMAN1990179}
J.~L. Elman, \enquote{Finding structure in time,} {\protect\JournalTitle{Cognitive Science}} \textbf{14}, 179--211 (1990).

\bibitem{10.1162/neco.1997.9.8.1735}
S.~Hochreiter and J.~Schmidhuber, \enquote{Long short-term memory,} {\protect\JournalTitle{Neural Comput.}} \textbf{9}, 1735–1780 (1997).

\bibitem{vaswani2017attention}
A.~Vaswani, N.~Shazeer, N.~Parmar, \emph{et~al.}, \enquote{Attention is all you need,} in \emph{Advances in neural information processing systems,}  (2017), pp. 5998--6008.

\bibitem{zhuang2020comprehensive}
F.~Zhuang, Z.~Qi, K.~Duan, \emph{et~al.}, \enquote{A comprehensive survey on transfer learning,}  (2020).

\bibitem{atmospheric_dispersion_corrector}
D.~qiang Su, P.~Jia, and G.~Liu, \enquote{Atmospheric dispersion corrector for the large sky area multi-object fibre spectroscopic telescope,} {\protect\JournalTitle{Monthly Notices of the Royal Astronomical Society}} \textbf{419}, 3406--3413 (2011).

\bibitem{xu2020real}
Y.~Xu, L.~Xin, J.~Wang, \emph{et~al.}, \enquote{A real-time automatic validation system for optical transients detected by gwac,} {\protect\JournalTitle{Publications of the Astronomical Society of the Pacific}} \textbf{132}, 054502 (2020).

\bibitem{danieli2020dragonfly}
S.~Danieli, D.~Lokhorst, J.~Zhang, \emph{et~al.}, \enquote{The dragonfly wide field survey. i. telescope, survey design, and data characterization,} {\protect\JournalTitle{The Astrophysical Journal}} \textbf{894}, 119 (2020).

\bibitem{liu2021sitian}
J.~Liu, R.~Soria, X.-F. Wu, \emph{et~al.}, \enquote{The sitian project,} {\protect\JournalTitle{Anais da Academia Brasileira de Ci{\^e}ncias}} \textbf{93} (2021).

\bibitem{law2022low}
N.~M. Law, H.~Corbett, N.~W. Galliher, \emph{et~al.}, \enquote{Low-cost access to the deep, high-cadence sky: the argus optical array,} {\protect\JournalTitle{Publications of the Astronomical Society of the Pacific}} \textbf{134}, 035003 (2022).

\bibitem{ofek2023large}
E.~Ofek, S.~Ben-Ami, D.~Polishook, \emph{et~al.}, \enquote{The large array survey telescope—system overview and performances,} {\protect\JournalTitle{Publications of the Astronomical Society of the Pacific}} \textbf{135}, 065001 (2023).

\bibitem{jia2023simulation}
P.~Jia, Q.~Jia, T.~Jiang, and Z.~Yang, \enquote{A simulation framework for telescope array and its application in distributed reinforcement learning-based scheduling of telescope arrays,} {\protect\JournalTitle{Astronomy and Computing}} p. 100732 (2023).

\bibitem{jia2023observation}
P.~Jia, Q.~Jia, T.~Jiang, and J.~Liu, \enquote{Observation strategy optimization for distributed telescope arrays with deep reinforcement learning,} {\protect\JournalTitle{The Astronomical Journal}} \textbf{165}, 233 (2023).

\end{thebibliography}






\end{document}